\documentclass[prd,aps,twocolumn,superscriptaddress,nofootinbib,showpacs,preprintnumbers,floatfix,balancelastpage,longbibliography]{revtex4-1}
\usepackage{graphicx, epsfig, bm, amsmath}
\usepackage{amssymb}
\usepackage{color}
\usepackage{float}
\usepackage{booktabs}
\usepackage{array}
\usepackage{makecell}
\usepackage[framemethod=TikZ]{mdframed}
\usepackage[colorlinks=true
,urlcolor=blue
,anchorcolor=blue
,citecolor=blue
,filecolor=blue
,linkcolor=red
,menucolor=blue
,linktocpage=true
,pdfproducer=medialab
,pdfa=true
]{hyperref}
\usepackage[english]{babel}
\usepackage[utf8]{inputenc}

\usepackage{wasysym}
\usepackage{setspace}
\usepackage{multirow}

\newcommand{\DM}{{\scriptscriptstyle \text{DM}}}
\newcommand{\aW}{\alpha_{\scriptscriptstyle W}}
\newcommand{\mW}{m_{\scriptscriptstyle W}}
\newcommand{\mZ}{m_{\scriptscriptstyle Z}}

\newcommand{\thetaW}{\theta_{\scriptscriptstyle W}}
\newcommand{\cW}{c_{\scriptscriptstyle W}}
\newcommand{\sW}{s_{\scriptscriptstyle W}}

\definecolor{armygreen}{rgb}{0.29, 0.33, 0.13}

\begin{document}


\preprint{IRFU-20-13, MIT-CTP 5120}

\vspace*{5pt}
\title{Prospects for Heavy WIMP Dark Matter with CTA: the Wino and Higgsino}

\author{Lucia Rinchiuso}
\affiliation{\!\!\mbox{ \footnotesize IRFU, CEA, D\'epartement de Physique des Particules, Universit{\'{e}} Paris-Saclay, F-91191 Gif-sur-Yvette, France}\,\,\,\vspace{0.7ex}}
\author{Oscar Macias}
\affiliation{\footnotesize Kavli Institute for the Physics and Mathematics of the Universe (WPI), University of Tokyo, Kashiwa, Chiba 277-8583, Japan\vspace{0.7ex}}
\affiliation{\footnotesize GRAPPA Institute, University of Amsterdam, 1098 XH Amsterdam, The Netherlands\vspace{0.7ex}}
\author{Emmanuel Moulin}
\affiliation{\!\!\mbox{ \footnotesize IRFU, CEA, D\'epartement de Physique des Particules, Universit{\'{e}} Paris-Saclay, F-91191 Gif-sur-Yvette, France}\,\,\,\vspace{0.7ex}}
\author{Nicholas L. Rodd}
\affiliation{\footnotesize Berkeley Center for Theoretical Physics, University of California, Berkeley, CA 94720, USA\vspace{0.7ex}}
\affiliation{\footnotesize Theoretical Physics Group, Lawrence Berkeley National Laboratory, Berkeley, CA 94720, USA\vspace{0.7ex}}
\author{Tracy R. Slatyer}
\affiliation{\footnotesize Center for Theoretical Physics, Massachusetts Institute of Technology, Cambridge, MA 02139, USA\vspace{0.7ex}}
\affiliation{\footnotesize School of Natural Sciences Institute for Advanced Study, Princeton, NJ 08540, USA\vspace{0.7ex}}

\begin{abstract}
\begin{centering}
\vspace{5pt}
{\bf Abstract}\\[4pt]
\end{centering}

TeV-scale particles that couple to the standard model through the weak force represent a compelling class of dark matter candidates.
The search for such Weakly Interacting Massive Particles (WIMPs) has already spanned multiple decades, and whilst it has yet to provide any definitive evidence for their existence, viable parameter space remains.
In this paper, we show that the upcoming Cherenkov Telescope Array (CTA) has significant sensitivity to uncharted parameter space at the TeV mass scale.
To do so, we focus on two prototypical dark matter candidates, the Wino and Higgsino.
Sensitivity forecasts for both models are performed including the irreducible background from misidentified cosmic rays, as well as a range of estimates for the Galactic emissions at TeV energies.
For each candidate, we find substantial expected improvements over existing bounds from current imaging atmospheric Cherenkov telescopes.
In detail, for the Wino we find a sensitivity improvement of roughly an order of magnitude in $\langle \sigma v \rangle$, whereas for the Higgsino we demonstrate that CTA has the potential to become the first experiment that has sensitivity to the thermal candidate.
Taken together, these enhanced sensitivities demonstrates the discovery potential for dark matter at CTA in the 1-100 TeV mass range.

\end{abstract}


\maketitle


\section{Introduction}
\label{sec:intro}

While dark matter (DM) is about 85\% of the total matter content of the universe, its fundamental nature is not known. An elementary particle with mass and couplings at the electroweak scale can naturally represent all the DM in the universe. 
Null searches at colliders and in direct detection have already excluded many realisations of this paradigm. 
Nonetheless, the WIMP (Weakly Interacting Massive Particle) miracle remains compelling, and especially for DM at the TeV mass range there remains uncharted parameter space.
Archetypal TeV DM WIMPs include the Wino and the Higgsino.
Both candidates could arise as the lightest supersymmetric particles, and can account for all the DM in significant parts of the supersymmetric parameter space, assuming a standard thermal history for the universe~\cite{Jungman:1995df}. 
More generally, Higgsino and Wino DM can be thought of as viable minimal extensions to the Standard Model (SM), independent of their high-scale origin.
That compelling DM candidates exist at the TeV mass scale provides a strong motivation to search for any hints of these new particles using ground-based imaging atmospheric Cherenkov telescopes (IACT), which have strong sensitivity in the TeV energy range.

In this work, we explore the sensitivity of the upcoming Cherenkov Telescope Array (CTA)~\cite{Consortium:2010bc} to Wino and Higgsino DM.
To do so, we use one of the most advanced calculations available for the gamma-ray annihilation spectra, where particularly for the case of the Wino there has been significant recent theoretical development.
Using the expected performances of the CTA observatory from the latest Monte Carlo simulations of the instrument response function (IRFs) for the Southern site~\cite{CTAperformances}, we compute the CTA sensitivity to Wino and Higgsino DM in the TeV to ten-TeV DM mass range using a 3-dimensional log-likelihood-ratio test statistic analysis.
There remains considerable uncertainty as to the relevant astrophysical backgrounds for very high-energy (VHE) gamma-ray searches in the Galactic Center (GC).
Accordingly, we will consider a wide range of possibilities for the background, and demonstrate that even in the most pessimistic scenarios the prospects for DM discovery at CTA are significant.

The paper is organized as follows. Section~\ref{sec:model} describes the modeling of the expected DM annihilation signals in gamma rays for the Wino and Higgsino DM. In Sec.~\ref{sec:background}, we discuss the astrophysical backgrounds in the GC region that are relevant for TeV DM searches. 
Sec.~\ref{sec:sensitivity} presents the expected  CTA performances,  the definition of the ROIs, the expected signal and background rates, and the statistical analysis method to compute the CTA sensitivity. The results are discussed in Sec.~\ref{sec:results} and we conclude in Sec.~\ref{sec:conclusion}.

\section{The Dark Matter Signal}
\label{sec:model}

In this work, we restrict our attention to two compelling DM benchmark candidates: the Wino and Higgsino.
The motivation for choosing these candidates is twofold.
First, they represent well-motivated DM candidates, that have proved challenging to test using other experimental techniques. 
Secondly, although they can be thought of as particularly simple extensions to the SM, the physics involved in determining the photon spectrum resulting from their annihilation is rich, and can be considered representative of the phenomena relevant for more general TeV-scale DM.
In this section we will briefly expand on each of these points.

Considering the problem from the top down, there are compelling theoretical motivations to believe that the Wino or Higgsino might be the DM of our universe.
Both candidates often arise generically as the lightest supersymmetric particle (LSP) in supersymmetric extensions of the SM~\cite{Giudice:1998xp,Randall:1998uk,ArkaniHamed:2006mb}.
Further, these candidates naturally arise as DM candidates near the weak scale in specific realizations such as split~\cite{ArkaniHamed:2004fb,Giudice:2004tc,ArkaniHamed:2004yi,Wells:2004di,Pierce:2004mk,Arvanitaki:2012ps,ArkaniHamed:2012gw,Fox:2014moa} or spread~\cite{Hall:2011jd,Hall:2012zp} supersymmetry.

Independent of high-scale motivations, the Wino and Higgsino can also be motivated from the bottom up, as they both represent simple TeV DM candidates, and can be taken as a testing ground for the type of theoretical effects that can be relevant for generic DM searches at CTA.
Indeed the electroweak triplet Wino has been identified as one of the simplest extensions to the SM to include a viable DM candidate, when viewed through the minimal DM lens~\cite{Cirelli:2005uq,Cirelli:2007xd,Cirelli:2008id,Cirelli:2009uv,Cirelli:2015bda}, and has been considered for almost three decades~\cite{Chardonnet:1993wd}.
The Higgsino can also arise in minimal models, as shown for example in Refs.~\cite{Mahbubani:2005pt,Kearney:2016rng}.
In passing we mention another particularly simple extension to the SM that would fall into this minimal class: the quintuplet.
Here the DM is a Majorana fermion transforming in the quintuplet representation of SU$(2)_W$; it represents an additional candidate of interest for CTA given its thermal mass is $14$ TeV~\cite{Mitridate:2017izz}, although we will not consider it in the present work.

Both the Wino and Higgsino couple through the SM, leaving only the masses of the DM and nearly degenerate states as free parameters in the theory.
The scale of the splitting between the various states, discussed further below, is significantly smaller than the overall mass scale.
For these reasons, assuming the particles are produced in the early universe as thermal relics, the overall mass scale is fixed by the observed DM density.
The required mass has been calculated as $2.9 \pm 0.1$ TeV for the Wino~\cite{Hisano:2006nn,Hryczuk:2010zi,Beneke:2016ync}, and $1.0 \pm 0.1$ TeV for the Higgsino~\cite{Hisano:2006nn,Cirelli:2007xd}.
These specific masses are of particular interest, and as we will show in this work are potentially within reach of CTA.
Nevertheless, it is worth exploring a larger mass range, which becomes plausible if we relax the assumption that the DM is a simple thermal relic produced within the standard cosmology.

Given their significant motivation, there have been a number of attempts to discover these forms of DM, for example see~\cite{Fan:2013faa,Cohen:2013ama,Krall:2017xij}.
At present, LHC searches have ruled out Wino masses below $\sim$$500$ GeV~\cite{Aaboud:2017mpt,Sirunyan:2018ldc}.
Unfortunately, the Higgsino is considerably more difficult to search for; even $\sim$$400$ GeV is a highly optimistic goal for the full LHC dataset~\cite{Low:2014cba}.
Further, thermal masses are out of reach for the LHC for both candidates, and potentially difficult to discover even at future 100 TeV colliders~\cite{Low:2014cba,Cirelli:2014dsa,Gori:2014oua,Berlin:2015aba}.
Discovery in direct detection is equally challenging.
The thermal Wino cross section is near the neutrino floor, and the Higgsino is even harder to probe, sitting below the floor~\cite{Hill:2011be,Hill:2013hoa,Hill:2014yka,Hill:2014yxa,Hisano:2015rsa}.
This leaves indirect detection, where the prospects for discovery at CTA will be explored in the present work.
The prospects for minimal supersymmetric DM candidates at CTA has previously been explored in~\cite{Roszkowski:2014iqa,Hryczuk:2019nql}, although we will approach the problem with updated predictions for the DM spectra and astrophysical background contributions.
Note also that indirect detection with anti-protons is an alternative search strategy, see for example~\cite{Ciafaloni:2012gs,Hryczuk:2014hpa,Cuoco:2017iax}, although the systematics associated with such searches are considerable.
{{Studies have been carried out in the framework of phenomenological supersymmetric extensions of the SM  (pMSSM) to demonstrate the complementarity of the different experimental techniques devised to detect DM depending on the composition of the LSP. As shown, for instance, in Refs.~\cite{Bergstrom:2010gh,Cahill-Rowley:2014boa,Roszkowski:2014iqa,Hryczuk:2019nql}, the CTA observations will open a unique discovery space for TeV scale WIMPs.}

Independent of its theoretical motivations, the physics involved in determining the Wino indirect detection signature has proven particularly rich.
The cross-section to line photons, which arises at one loop, was first calculated more than twenty years ago~\cite{Bergstrom:1997fh,Bern:1997ng,Ullio:1997ke}.
This result is not sufficient to obtain all relevant $\mathcal{O}(1)$ contributions to both the rate for Wino annihilations and the resulting photon spectrum.
To do so there are four effects that must be included.
\begin{enumerate}
\item The Sommerfeld enhancement, where a significant correction to the cross section arises from the potential generated by the exchange of electroweak particles between the Wino states~\cite{Hisano:2003ec,Hisano:2004ds,Cirelli:2007xd,ArkaniHamed:2008qn,Blum:2016nrz};
\item Continuum emission of photons with $E \ll m_{\DM}$ resulting from the decay of final state $W$ and $Z$ bosons, the spectrum of which can be determined using for example PPPC 4 DM ID~\cite{Cirelli:2010xx};
\item Resummation of Sudakov double logarithms of the form $\aW \ln^2(m_{\DM}/\mW)$, which become significant when the DM mass is well above the scale of the electroweak particles which mediate the annihilation~\cite{Hryczuk:2011vi,Bauer:2014ula,Ovanesyan:2014fwa,Baumgart:2014saa,Baumgart:2015bpa,Ovanesyan:2016vkk};
\item Inclusion of endpoint photons, which have $E = z m_{\DM}$ with $1-z \ll 1$.
At any instrument with finite energy resolution, such as CTA, these photons can become indistinguishable from the line associated with the two body final state where $z=1$.
Due to the phase space restriction on photons most likely to be mixed with the line, their contribution is enhanced as $\aW \ln^2(1-z)$, and again these logs must be resummed~\cite{Baumgart:2015bpa,Baumgart:2017nsr,Baumgart:2018yed}.\footnote{For annihilation of neutralinos with no Sommerfeld enhancement, these contributions have previously been calculated at fixed order in~\cite{Bergstrom:2005ss,Bringmann:2007nk}; in the presence of Sommerfeld enhancement, there will be additional endpoint contributions arising from the conversion of neutralinos into a virtual chargino pair through the long-range potential, followed by annihilation of the charginos.}
\end{enumerate}

The first calculation involving all of these effects for the Wino was performed in Ref.~\cite{Baumgart:2017nsr}, and has now been extended to next-to-leading logarithmic (NLL) accuracy~\cite{Baumgart:2018yed}.
In the present work we will make use of the full NLL spectrum from that reference, and in many ways that aspect of our analysis will represent an update of the H.E.S.S. Wino sensitivity estimates performed in~\cite{Rinchiuso:2018ajn} for the case of CTA.
We note that the calculation in~\cite{Baumgart:2018yed} was performed under the assumption that $\mW/(2m_{\DM})$ is much smaller than the instrumental energy resolution, $\Delta E/E$.
Given that for CTA $\Delta E/E \sim 10\%$, we estimate that the predicted spectrum begins to become unreliable for $m_{\DM} \lesssim 1$ TeV.
Fixing this would require matching our predictions onto a calculation valid in this regime, such as the calculation in Refs.~\cite{Beneke:2018ssm,Beneke:2019vhz}.
In this work we will show limits down to $m_{\DM} = 600$ GeV, and given the above caveat we emphasize that results in this low mass region are subject to larger theoretical uncertainties.

In the case of the Higgsino, a full calculation involving all effects relevant for CTA has not yet been performed, and is well beyond the scope of the present analysis.
A number of results do exist, however, including the full Sommerfeld calculation~\cite{Hisano:2004ds}.
In addition, the work in~\cite{Baumgart:2015bpa} demonstrated that the resummed endpoint contribution for the Higgsino is likely to be crucial, and to lead to a large $\mathcal{O}(1)$ correction. 
Most significantly, a full NLL' calculation of the effects in a framework assuming the energy resolution is of order $\mW/m_{\DM}$ or $(\mW/m_{\DM})^2$ has recently been performed in~\cite{Beneke:2019gtg}, although for CTA this assumption will not hold across the entire mass range considered here.
Given this state of affairs, in the present work we will use just the tree-level annihilation rate, supplemented with the Sommerfeld enhancement, to produce both line photons and continuum emission (i.e. photons arising from decays of the primary annihilation products).
As we will show, within this simplified framework the prospects for the Higgsino at CTA appear encouraging.
Nevertheless, we caution that we are using theoretical predictions that are known to be missing large $\mathcal{O}(1)$ corrections, a caveat that applies to all Higgsino results shown in this work.
That our results indicate CTA may well be able to probe the thermal Higgsino only reinforces the need for the full calculation to be performed.

In the Higgsino case, there are two additional parameters which must be specified beyond the DM mass, namely the splittings between the charged and neutral states, $\delta m_+$ and $\delta m_N$.
In the Wino case this is not an additional degree of freedom, as when all the other superpartners are much heavier, the splitting is dominated by radiative effects, and has been calculated at two-loops to be $m_{\chi^{\pm}}-m_{\chi^0} \simeq 164.4$ MeV~\cite{Ibe:2012sx}.
If the splittings were purely radiative in the Higgsino case, the neutral states would be of equal mass ($\delta m_N =0$) and would both contribute to the DM, allowing for tree-level scattering between DM and visible particles via Z exchange.
This scenario is strongly excluded by constraints from direct detection; evading this limit requires the heavier neutral state to be kinematically inaccessible in direct-detection experiments, suggesting $\delta m_N \gtrsim 200$ keV for TeV-scale DM.
Such small splittings can be easily induced in supersymmetric scenarios by a tiny mixing of the Higgsino with the heavier neutralinos.
For the Higgsino, consequently, there is a wide space of possible mass splittings.
Given that our Higgsino spectrum is representative and not exact, as mentioned above, we will not attempt an exhaustive scan of the allowed model space.
Instead, we take two representative values following~\cite{Baumgart:2015bpa}.
Specifically, we consider
\begin{itemize}
\item \textit{Scenario 1:} $\delta m_N = 200$ keV and $\delta m_+ = 350$ MeV. 
This scenario represents the case where the neutral mass splitting saturates the direct detection bound and the charged mass splitting is set to its radiative value. 
\item \textit{Scenario 2:} $\delta m_N = 2$ GeV and $\delta m_+ = 480$ MeV.
This scenario is chosen to contrast with the above, so that now the ratio of the splittings has been inverted, and we have $\delta m_N \gg \delta m_+$.
A $\delta m_N$ splitting at this level can be generated in the Minimal Supersymmetric Standard Model by gauginos just a factor of few heavier than the Higgsino; it thus represents the upper end of the splittings expected from mixing effects.
\end{itemize}
All results for the Higgsino will be shown for both scenarios.

\subsection{Dark Matter Density Distribution in the Galactic Center}

In addition to the details of the particle nature of dark matter, the GC annihilation signal depends critically on the distribution of DM around the centre of the Milky Way.
Yet at present, the distribution of the density of DM in this region of the galaxy is neither firmly predicted from simulations nor significantly constrained by observations.
N-body simulations containing only DM motivate density profiles rising steeply toward the GC, and for our fiducial model we choose one parameterization of these observations, the Einasto profile~\cite{1965TrAlm...5...87E}
\begin{equation} 
\rho_\text{Einasto} = \rho_0 \exp\left[-\frac{2}{\alpha} \left( \left(\frac{r}{r_s}\right)^\alpha - 1 \right)\right],
\end{equation}
where $r$ is the Galactocentric radius. 
Our fiducial model is defined by the following parameter for the Milky Way Galaxy we take $\alpha = 0.17$ \cite{Abramowski:2011hc}, $r_s=20$ kpc~\cite{Pieri:2009je}, and $\rho_0$ chosen so that $\rho_\DM(r_{\odot}) = \rho_{\odot} = 0.39$ GeV/cm$^3$~\cite{Catena:2009mf}, where $r_{\odot} = 8.5 \text{ kpc}$ is the estimated distance from the Sun to the GC.
These values are motivated by a comparison to earlier results, however we note that the exact value for a number of these parameters are currently being refined.
For example, more recent measurements have found that $r_{\odot} = 8.127 \text{ kpc}$ \cite{Abuter:2018drb}.
Improvement to our understanding of the local DM density are being pursued on a number of fronts, see for example~\cite{Read:2014qva} for a review.
We emphasize that any change to $\rho_{\odot}$ can be trivially propagated to our results by rescaling the predicted signal by a factor of $[\rho_{\odot}/(0.39~{\rm GeV/cm^3})]^2$.

One systematic uncertainty associated with the GC DM density that can have a significant impact on the present analysis is the possibility that true density is reduced by to the presence of a core in the inner galaxy.
The incorporation of baryonic matter and its associated feedback into N-body simulations have demonstrated that these effects can flatten out the DM density distribution at small $r$, producing a constant-density ``core''.
For Milky-Way-sized galaxies, the core radius can be of order 1 kpc~\cite{Chan:2015tna}, or even larger; depending on the modeling of baryonic physics, cores extending to $\sim5$ kpc can potentially be obtained~\cite{Mollitor:2014ara}.

At present, the exact size of such a core is highly uncertain.
To account for the uncertainty in the DM distribution at small Galactocentric radii, and the possibility of kpc-scale cores in the region of interest (ROI), we study the gamma-ray signals associated with a cored density profile.
We empirically parameterize constant-density cores by setting the density profile to $\rho_\text{Einasto}(r)$ for $r > r_\text{\rm c}$, and to $\rho_\DM(r_\text{c}) = \rho_{\rm Einasto}(r_\text{\rm c})$ for $r < r_\text{\rm c}$.
We fix the normalization of the density profile at $r=r_{\odot}$, i.e. $\rho_\DM(r_{\odot}) = \rho_\odot$ in all cases.

\subsection{Annihilation Signal Spectrum}

The total photon flux observed from DM annihilation, in a given ROI, is given by
\begin{equation}
\label{eq:flux}
\frac{d \Phi_\gamma}{d E}
=
\frac{\langle \sigma v \rangle_{\rm line}}{8\pi m_{\DM}^2} \frac{d N_{\gamma}}{d E} \times J\,,
\end{equation}
where the astrophysical factor, or $J$-factor, is given by
\begin{equation}
\label{eq:jfactor}
J = \int_{\rm ROI} d \Omega\, \int ds\, \rho^2_{\DM}\,.
\end{equation}

We emphasize at the outset that the use of $\langle \sigma v \rangle_{\rm line}$ within Eq.~(\ref{eq:flux}) does not imply we are only considering the annihilation of dark matter to two body final states producing a photon at almost exactly $m_{\DM}$.
There is an inherent freedom to redefine what means in that equation by $\langle \sigma v \rangle$ and the cross section per annihilation, $dN_{\gamma}/dE$, as long as the product is left unchanged.
We have exploited this freedom to write the cross section in a convenient form, as the cross section to produce two photons,
\begin{equation}
\langle \sigma v \rangle_{\rm line} = \langle \sigma v \rangle_{\gamma \gamma + \gamma Z/2}\,.
\end{equation}
In detail $\sigma_{\rm line}$ corresponds to the cross section for ${\rm DM}\,{\rm DM} \to \gamma \gamma$ plus half the cross section for ${\rm DM}\,{\rm DM} \to \gamma Z$, as there is only a single photon in the latter process.
This is a convenient choice, as this cross section is traditionally how limits on the Wino and Higgsino are presented.
Using this definition, if the spectrum consisted only of the exclusive line, then it would simply be given by $2 \delta(E-m_{\DM})$.
Yet as emphasized, even though we will use $\langle \sigma v \rangle_{\rm line}$ to parameterize our rate, it is not the only final state we include.
More generally, we have
\begin{equation}
\label{eq:dNdE}
\frac{dN_{\gamma}}{dE} = 2 \delta (E-m_{\DM})
+ \frac{dN^{\rm ep}_{\gamma}}{dE} 
+ \frac{dN^{\rm ct}_{\gamma}}{dE}\,, 
\end{equation}
where the normalizations for the endpoint (ep) and continuum (ct) spectra are determined relative to the line cross section.
A more detailed discussion of this point, and the detailed form the endpoint and continuum spectra appearing in Eq.~(\ref{eq:dNdE}), can be found in~\cite{Baumgart:2017nsr}.

As discussed above, in the case of the Wino we will use the full analytic NLL calculation of the endpoint spectrum and $\langle \sigma v \rangle_{\rm line}$ provided in~\cite{Baumgart:2018yed}.
The continuum spectrum from production of gauge bosons and their subsequent decay is calculated as described in that work.

For the Higgsino, no endpoint contribution is included, and the line and continuum cross sections are estimated from a tree-level calculation including the non-perturbative effects of Sommerfeld enhancement.
The Sommerfeld-enhanced cross section for each final state channel $X$ is determined as:
\begin{equation} 
(\sigma v_\text{rel})_{\chi^0 \chi^0 \rightarrow X} = 2 \sum_{j j^\prime} s_{0j} (\Gamma_X)_{j j^\prime} s_{0j^\prime}^*, 
\end{equation}
where $\Gamma_X$ is a channel-specific ``annihilation matrix'', and the $s_{0j}$ coefficients describe the Sommerfeld vector appropriate to the $\chi^0 \chi^0$ initial state, derived by solving the Schr\"{o}dinger equation, as discussed in Ref.~\cite{Cohen:2013ama} (and following the notation in that work).
The prefactor of 2 accounts for the fact that our initial state consists of two identical DM particles. The potential matrix $V(r)$ for the Schr\"{o}dinger equation is given by \cite{Hisano:2004ds}:
\begin{align} 
\begin{pmatrix} 2 \delta m_+ - \frac{\alpha}{r} - \frac{\aW x^2}{4 \cW^2} \frac{e^{-\mZ r}}{r} 
& -\frac{\sqrt{2} \aW e^{-\mW r}}{4 r} 
&  -\frac{\sqrt{2} \aW e^{-\mW r}}{4 r} \\  
-\frac{\sqrt{2} \aW e^{-\mW r}}{4 r}
& 2 \delta m_N
& -\frac{\aW e^{-\mZ r}}{4 \cW^2 r} \\  
-\frac{\sqrt{2} \aW e^{-\mW r}}{4 r} 
& -\frac{\aW e^{-\mZ r}}{4 \cW^2 r} 
& 0 \end{pmatrix}.
\end{align}
The first, second, and third rows/columns correspond respectively to the $\chi^+ \chi^-$ two-particle state, the two-particle state comprised of the heavier neutral species $\chi^1 \chi^1$, and the two-particle DM-DM state $\chi^0 \chi^0$.
Here $x = 1 - 2 \sW^2$, $\cW = \cos\thetaW$, and $\sW=\sin\thetaW$. 

The tree-level annihilation matrices appropriate to the Higgsino are given by \cite{Hisano:2004ds}:\footnote{We have corrected the off-diagonal terms for the annihilation matrix in Ref.~\cite{Hisano:2004ds}, bringing the inclusive annihilation rates into agreement with Ref.~\cite{Cirelli:2007xd}.}
\begin{align} 
\Gamma_{W^+W^-} & = \frac{\pi \aW^2}{64 m_{\DM}^2} \begin{pmatrix} 8 & 4 \sqrt{2} & 4 \sqrt{2} \\ 4 \sqrt{2} & 4 & 4 \\ 4 \sqrt{2} & 4 & 4 \end{pmatrix}, \nonumber \\
 \Gamma_{Z^0 Z^0} & = \frac{\pi \aW^2}{64 \cW^4 m_{\DM}^2} \begin{pmatrix} 4 x^4 & 2 \sqrt{2} x^2& 2 \sqrt{2} x^2 \\ 2 \sqrt{2} x^2 & 2 & 2 \\ 2 \sqrt{2} x^2 & 2 & 2 \end{pmatrix}, \nonumber
\end{align}
\begin{align}
\Gamma_{\gamma Z^0} & =  \frac{\pi \alpha \aW}{2 \cW^2 m_{\DM}^2} \begin{pmatrix} x^2 & 0 & 0 \\ 0 & 0 & 0 \\ 0 & 0 & 0 \end{pmatrix}, \nonumber \\
 \Gamma_{\gamma \gamma} & =  \frac{\pi \alpha^2}{m_{\DM}^2} \begin{pmatrix} 1 & 0 & 0 \\ 0 & 0 & 0 \\ 0 & 0 & 0 \end{pmatrix},
 \end{align}
 where $\alpha = \alpha_{\scriptscriptstyle \text{EM}}$.

In both cases the continuum spectrum arising from $W$ and $Z$ decays is determined using~\cite{Cirelli:2010xx} (note that masses above 100 TeV would require manual generation of the spectra, e.g. using \textsc{Pythia} 8.215~\cite{Sjostrand:2006za,Sjostrand:2007gs,Sjostrand:2014zea}, but we do not consider such high masses in this work).

\section{Backgrounds in the GC region}
\label{sec:background}
\begin{figure*}[htbp] 
 \begin{minipage}{\textwidth}
\begin{center}
\includegraphics[width=0.45\textwidth]{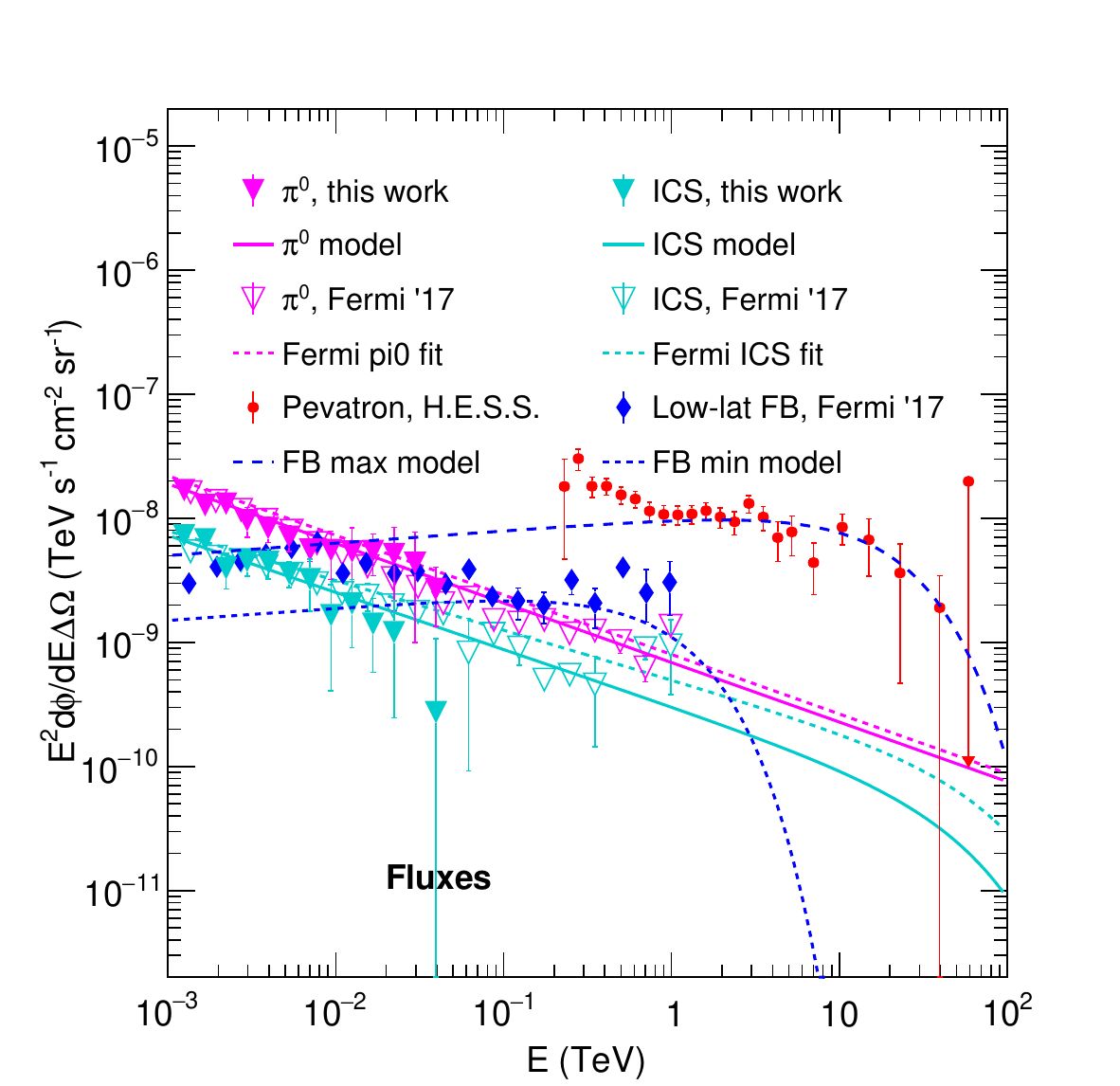}
\includegraphics[width=0.45\textwidth]{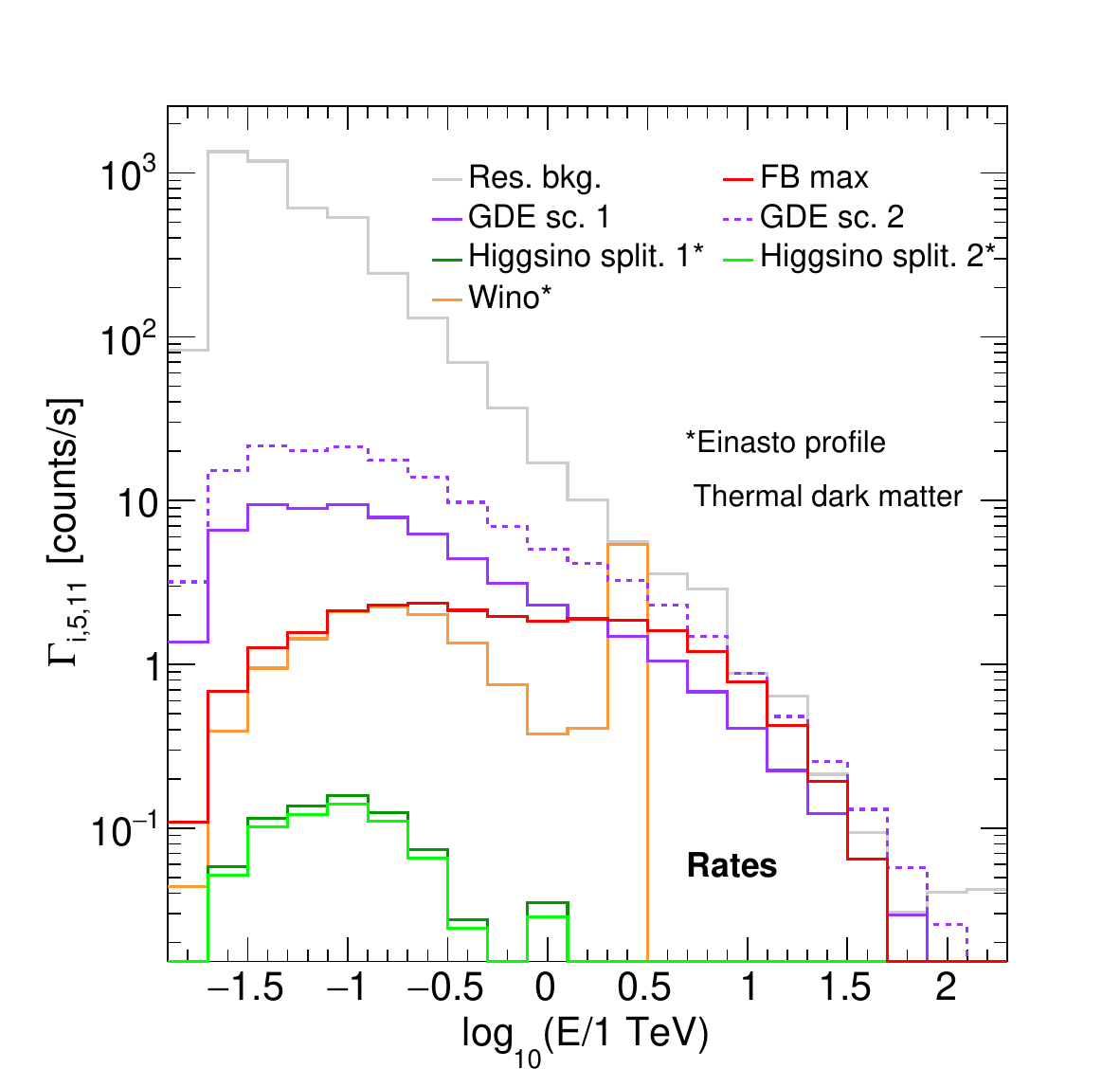}
\caption{{\it Left panel:} VHE astrophysical emissions in the GC region. The energy differential fluxes are plotted for 
the $\pi^0$ (pink open  triangles) and ICS (cyan open triangles) components of the Galactic Diffuse Emission measured by {\it Fermi}-LAT extracted from Ref.~\cite{TheFermi-LAT:2017vmf}, and for the $\pi^0$ (pink filled triangles) and ICS (cyan filled triangles) components derived from an alternative analysis~\cite{Macias:2016nev} performed in the inner $15^\circ\times15^\circ$ of the GC. The filled and open triangles are used for fitting a power-law or a power-law with exponential cut-off, and then we use the fitted curves to normalize the maps in the CTA energy range. The H.E.S.S. Pevatron spectrum (red dots) and the low-latitude spectrum of the Fermi Bubbles (blue filled squares)~\cite{Abramowski:2016mir} are also shown. Conservative (dashed blue line) and optimistic (log-dashed blue line) parametrizations of the Fermi Bubbles component are plotted. Note that the region corresponding to the Galactic ridge emission is excluded from the region of interest (see Sec. IV.B).
{\it Right panel:} Expected differential count rate as a function of energy for the expected signal and backgrounds in
the ROI (5,11), a 0.5$^\circ\times$0.5$^\circ$ squared pixel centered at $l=2.75^\circ,b=-0.25^\circ$ (see Sec. IV.B for more details on the ROI definition). The expected DM rates are plotted for a 3 TeV Wino with an annihilation cross section of $2.3\times10^{-26}$~cm$^3$s$^{-1}$ (orange solid line), and for a 1 TeV Higgsino with an annihilation cross section of $1.1\times 10^{-28}$~cm$^3$s$^{-1}$ ($9.2\times 10^{-29}$~cm$^3$s$^{-1}$) (green solid lines) assuming $\delta m_N=200$~keV ($\delta m_N=2$~GeV) and $\delta m_+=480$~MeV ($\delta m_+=350$~MeV). 
For the background we show the residual background, the backgrounds for GDE scenarios 1 and 2, and the maximum Fermi bubbles contribution. The DM signal curves correspond to the full spectra, including the line component, and have been convolved with the CTA energy resolution.}
\label{fig:fluxes}
\end{center}
 \end{minipage}
\end{figure*}

\subsection{Cosmic-ray background}
\label{sec:CRs}
The main background for the DM search with IACTs comes from hadron and electron cosmic rays (CRs) that are misidentified as gamma rays.
The numerous interactions of CR hadrons (protons and nuclei) in the Earth's atmosphere give rise to hadronic showers which may induce electromagnetic subcascades due to the decay of neutral pions produced in inelastic CR interactions. 
While the hadronic and electromagnetic showers can be efficiently discriminated through shape analysis of the shower image~\cite{2009APh32231D,Parsons:2014voa} and,  possibly, the arrival time of the shower front~\cite{Aharonian:2008zza}, a fraction of the hadronic cosmic rays cannot be distinguished from photons due to the finite rejection power of the instrument, and
the electron-induced showers are very similar to the gamma-ray ones~\cite{Aharonian:2008aa}. Misidentified hadrons and electrons\footnote{Local ($\lesssim$ 1 kpc) CR electron/positron sources may leave an imprint in the arrival directions of VHE electrons. However, no anisotropy has been detected so far~\cite{Abdollahi:2017kyf}.} produce an irreducible isotropic background, referred to hereafter as the residual background. The residual background for CTA is computed from accurate simulations of the incoming CR fluxes including protons and heavier nuclei, described by power-law spectra, as well as electrons and positrons, described by a log-normal peak on top of a power-law.  The publicly-available IRFs of CTA used in the present work are provided in Ref.~\cite{CTAperformances}. The source file {\it CTA-Performance-prod3b-v1-South-20deg-average-50h.root} is used to extract the energy-dependent effective area, background rate and energy resolution. The IRFs are taken for the Southern site of CTA, which is most relevant for observations of the inner region of the Milky Way, at a 20$^{\circ}$ mean zenith angle (which we expect to approximate the average zenith angle for observations of the GC). 

\subsection{Definition of the Regions of Interest}
The survey of the inner Galactic halo is one of the key-science observation programmes of CTA~\cite{Acharya:2017ttl}. 
The central survey region plans a deep exposure of more than 500 h expected in the inner 5$^\circ$ of the GC with, in addition, 300 h of exposure from the extended survey to cover regions out 10$^\circ$ from the Galactic plane.
The ROI for the annihilating DM search is defined as a square region in Galactic coordinates of 10$^\circ$ side length, centered at the GC.  
Defining several sub-ROIs improves the sensitivity to DM by exploiting the features of the spatial behavior of the expected DM signal with respect to backgrounds. The search region is split into 400 square 0.5$^\circ \times$0.5$^\circ$ pixels. No significant impact of the choice of the spatial binning size is noticed given the expected photon statistics obtained in each bin from the CTA inner Galactic halo survey. Varying the bin size (within a range that gives reasonable photon statistics per bin) has been tested and has a negligible impact on the results. 
The solid angle of the spatial bin $jk$ is given by:
\begin{equation}
\displaystyle \Delta\Omega_{\rm jk}=\int_{\Delta l}\int_{\Delta b}{\rm d}b_{\rm k}{\rm d}l_{\rm j}\cos b_{\rm k},
\end{equation}
where $l$ and $b$ are the Galactic longitude and latitude, respectively, given by $l = l_{\min} + j \Delta l$ and $b = b_{\rm min} + k \Delta b$ with $b_{\rm min} = -4.75^\circ$ and $l_{\rm min} = -4.75^\circ$.
$\Delta b=0.5^\circ$ and $\Delta\l=0.5^\circ$ are the sizes of the square pixel in longitude and latitude, respectively. 

Following Refs.~\cite{Abdallah:2018qtu, Abdallah:2016ygi}, the region of $\pm0.3^\circ$ around the Galactic plane is excluded as being dominated by standard astrophysical sources of VHE gamma rays. In addition a disk of radius 0.4$^\circ$ is discarded at the position of HESS J1745-303, one of the brightest extended VHE gamma-ray sources in the overall ROI. In addition, circular regions of $0.25^\circ$ radius centered on the selected {\it Fermi}-LAT source nominal positions are excluded.

\subsection{TeV Diffuse Emission in the Galactic Center}
\label{sec:TeVdiffuse}
The GC is a very crowded region where significant VHE gamma-ray emission arises from various astrophysical objects and production processes. In addition to pointlike sources such as HESS J1745-290~\cite{Aharonian:2004wa,Aharonian:2009zk} spatially coincident with the supermassive black hole Sagittarius A* lying at the gravitational center of the Milky Way, diffuse emission will also contribute to the total gamma-ray flux. Deep observations of the GC region carried out by H.E.S.S. reveal the detection of VHE emission correlated with massive clouds of the Central Molecular Zone~\cite{Aharonian:2006au}, and more recently extended emission in the inner 50 pc of the GC~\cite{Abramowski:2016mir}, from PeV protons interacting in the interstellar medium. 

At lower energies, the Galactic diffuse emission (GDE) constitutes about 80\% of all the photons detected by \textit{Fermi}-LAT in the energy range of a few MeV to $\sim 1$ TeV~\cite{Ackermann:2012}. The GDE results from the interactions of energetic CR particles with interstellar material and ambient photons,
possibly including individual diffuse sources.\footnote{In what follows, the GDE model restricts to the CR-induced interstellar emission model since the main diffuse sources relevant for the analysis here are masked in the region of interest.}
 The main processes giving rise to the GDE are $\pi^0$-decay, Bremsstrahlung, and Inverse Compton (IC) scattering. In the \textit{Fermi}-LAT energy band, current efforts to detect DM in this sky region are limited by uncertainties in the models for these three components; this is not the case for current H.E.S.S. VHE gamma-ray searches, but the greater sensitivity of CTA will likely render the GDE contribution important even in the VHE regime. Thus unraveling a potential DM signal from the GC observations by CTA, or setting robust constraints, will require the construction of GDE models that are as realistic as possible. 

The \textit{Fermi}-LAT collaboration has developed a GDE model which is publicly available and is the standard in most \textit{Fermi} analyses~\cite{Acero:2016qlg}. This model was constructed with a data-driven approach in which the $\pi^0$-decay and Bremsstrahlung gamma rays were modeled as a linear combination of spatial templates describing the distribution of interstellar gas in the Galaxy. For the IC emission this work used the CR propagation code GALPROP~\cite{Galprop}. In order to account for some extended positive gamma-ray residuals that have been detected in various regions of the sky, the \textit{Fermi}-LAT collaboration also included several empirical maps in their GDE model. Among them, and relevant to this work, are the Fermi Bubbles~\cite{Su:etal2010,FermiLat:Bubbles} (see Sec.~\ref{sec:FBs} for details). 

However, a limitation of the \textit{Fermi} GDE model is that it only allows for its overall normalization to be varied in the fits to the gamma-ray data. This is a very good approximation for analyses of compact astrophysical objects (also called gamma-ray point sources) but can have an impact in studies of extended objects like that of a putative DM source.

The systematic uncertainties in the GDE model can also be explored using GALPROP. Ackermann et al. (2012)~\cite{Ackermann:2012} performed fits to the gamma-ray data using a grid of alternative GDE models created with different propagation parameter setups,  CR halo sizes and 2D CR source distributions. One of the advantages of this approach is that each component of the GDE can be included in the maximum likelihood fits with independent normalizations. 

Interestingly, the most recent release of GALPROP (v56)~\cite{Porter:2017vaa} contains more realistic 3D models for the Interstellar Radiation Field (ISRF) and interstellar gas distributions. Compared to older versions of the code, the new models abandon the common assumption of 2D Galactocentric cylindrical symmetry in the propagation of CRs, which is expected to have important effects for analyses of the GC region. 

In the present study, we make use of GALPROP v56 to reproduce a single representative model for the GDE taken from Ref.~\cite{Porter:2017vaa}, and include it in our gamma-ray pipeline. For the chosen model, the cosmic rays that source the observed diffuse emission follow a distribution that is half 2D disc and half 3D spiral arms. The spiral arms template is based on~\cite{Freudenreich:1997bx}. For the specific details of this model we refer to~\cite{Porter:2017vaa}, and see also~\cite{Macias:2019omb, Abazajian:2020tww} where the model used is referred to as ``F98-SA50". Going forward, we refer to this specific GDE model as ``GDE scenario 1''. The maps of this GDE model in Galactic coordinates are given in the top panels of Fig.~\ref{fig:maps1TeV} in terms of integrated flux in the energy bin centered at 1~TeV and a spatial pixel of size $0.5^\circ\times0.5^\circ$.
We use a data-driven prescription for the energy spectrum of this GDE model. For energies below 50 GeV, we first fit the spatial maps produced by GALPROP in each energy bin to the \emph{Fermi}-LAT data in a $15^\circ \times 15^\circ$ region centered on the GC, separately floating the gas-correlated and ICS components~\cite{Macias:2016nev}. The resulting fluxes associated with each component are shown in the left panel of Fig.~\ref{fig:fluxes} (filled pink and cyan triangles). We then use these data points to find simple parametric descriptions of the spectra for the two components; it is these parametric forms that are used in the remainder of the analysis.

Specifically, we fit a power-law to the derived data points for the spectrum of the gas-correlated emission (filled pink triangles in Fig.~\ref{fig:fluxes}), and a power-law with exponential cutoff to the derived data points for the spectrum of the ICS (filled cyan triangles in Fig.~\ref{fig:fluxes}). The resulting best-fit values for the spectral index $\Gamma$, normalization $\phi_0$ and energy of cutoff $E_\text{cut}$ are given in Tab.~\ref{tab:GDEmodel}. The spatial maps produced by GALPROP, at each energy, are then re-normalized so that their total fluxes in the region of interest are described by these spectra. Since our purpose is to estimate the likely contribution of the GDE at higher energies, we prefer this approach to simply fitting the maps in each energy bin to the \emph{Fermi}-LAT data, because at high energies these data become noisy (due to low photon statistics) and the error bars on the GDE components can be very large.

\begin{figure*}[htbp] 
\begin{center}
\includegraphics[width=0.45\textwidth]{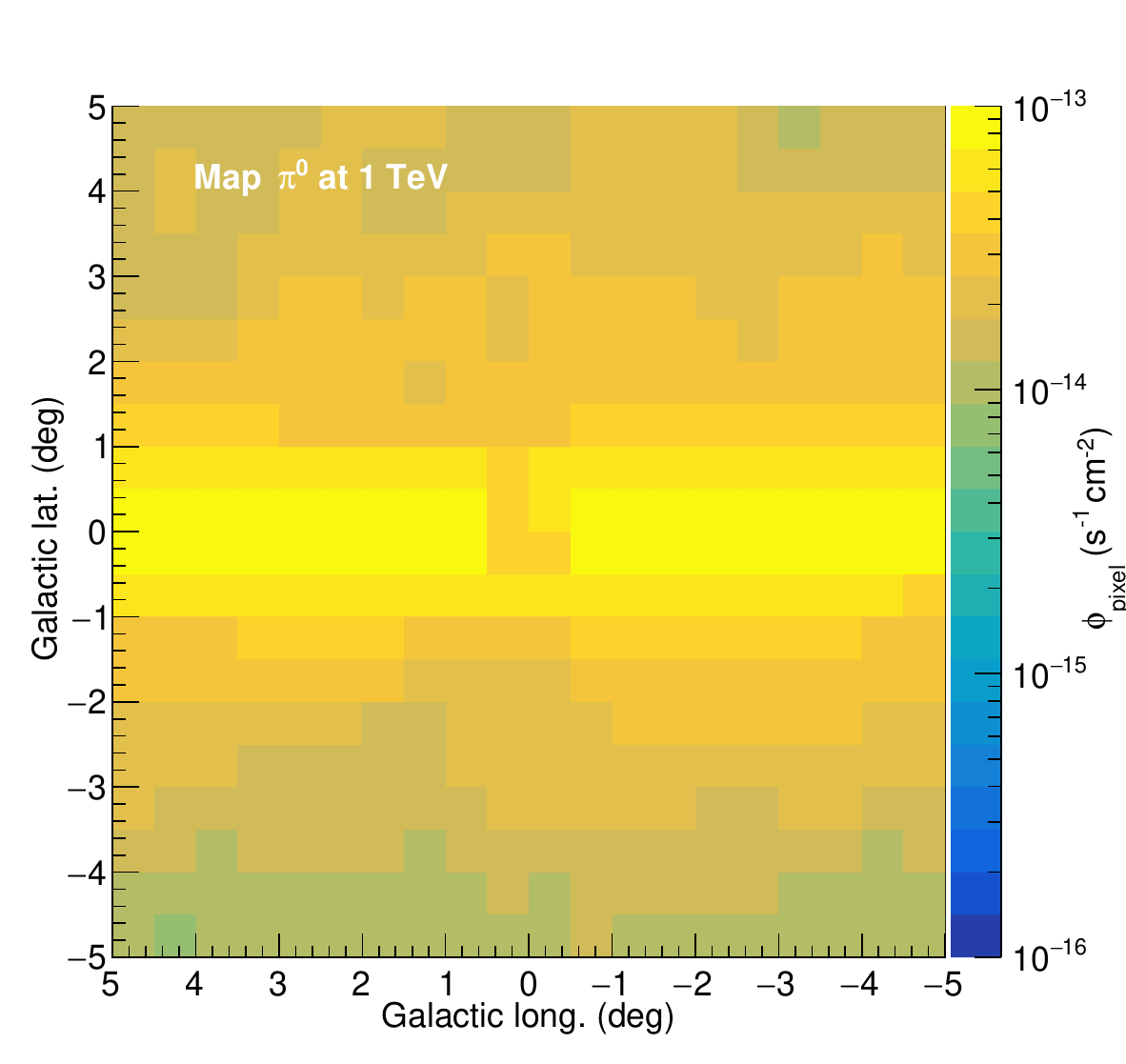}
\includegraphics[width=0.45\textwidth]{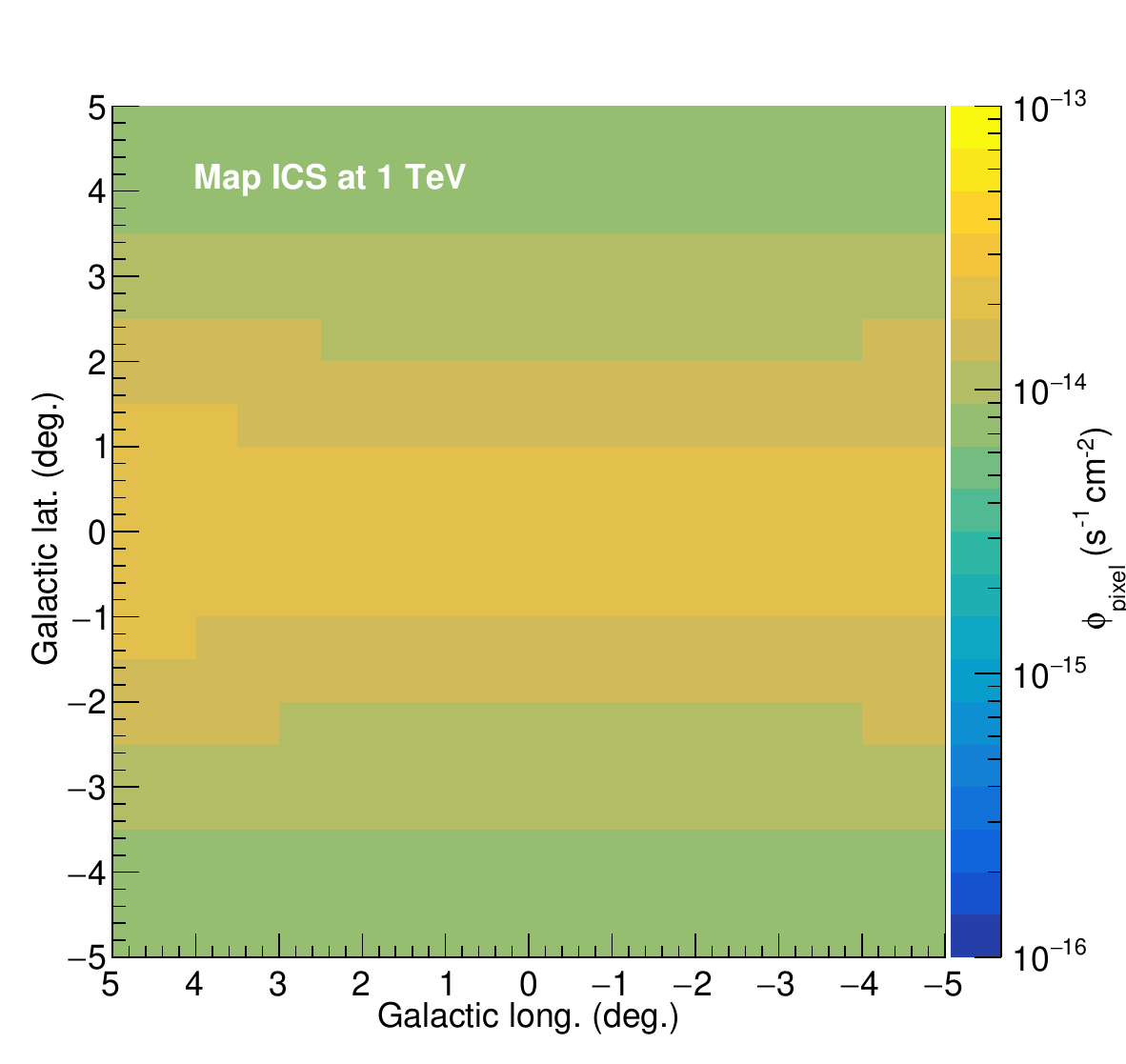}
\includegraphics[width=0.45\textwidth]{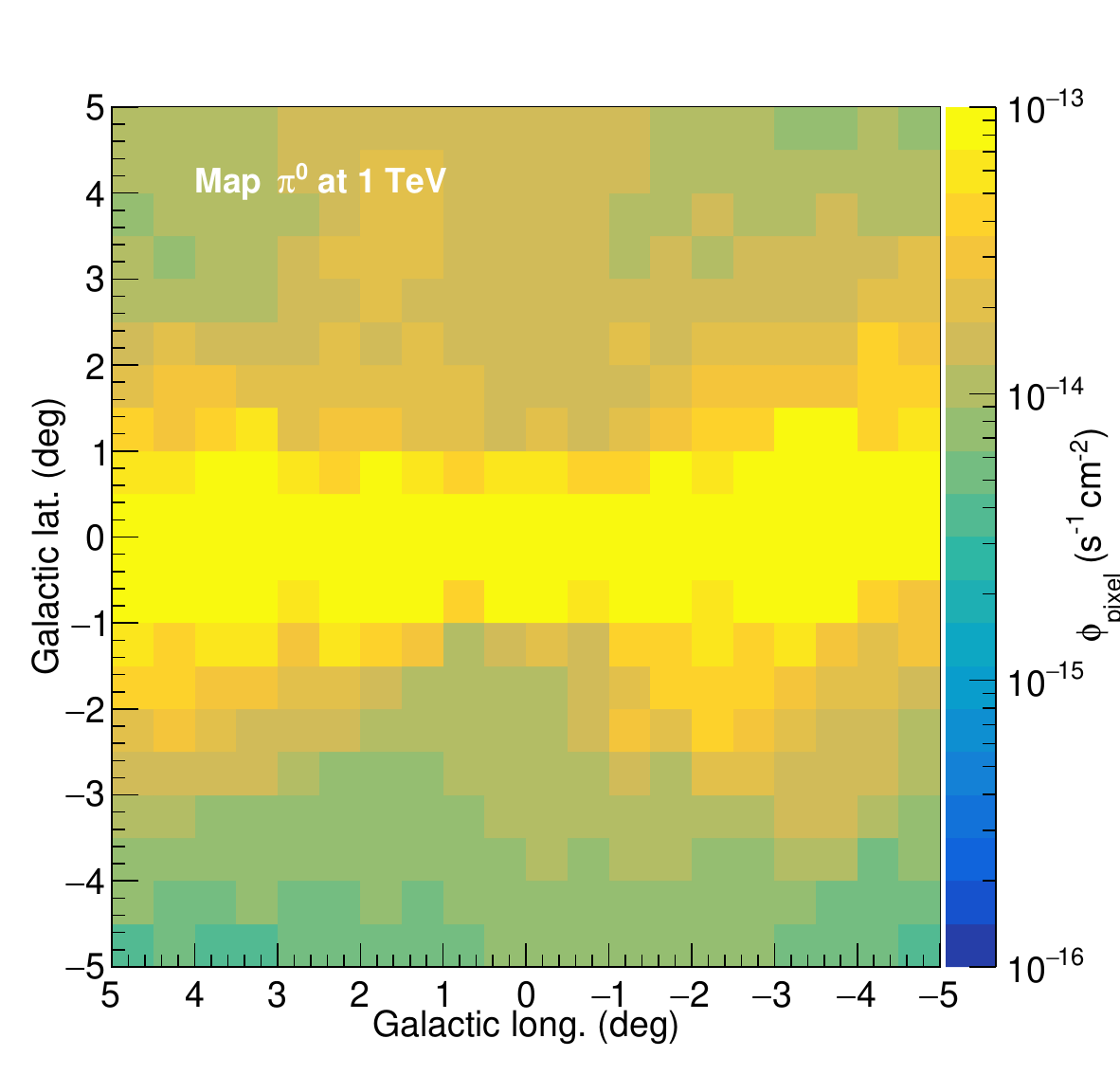}
\includegraphics[width=0.45\textwidth]{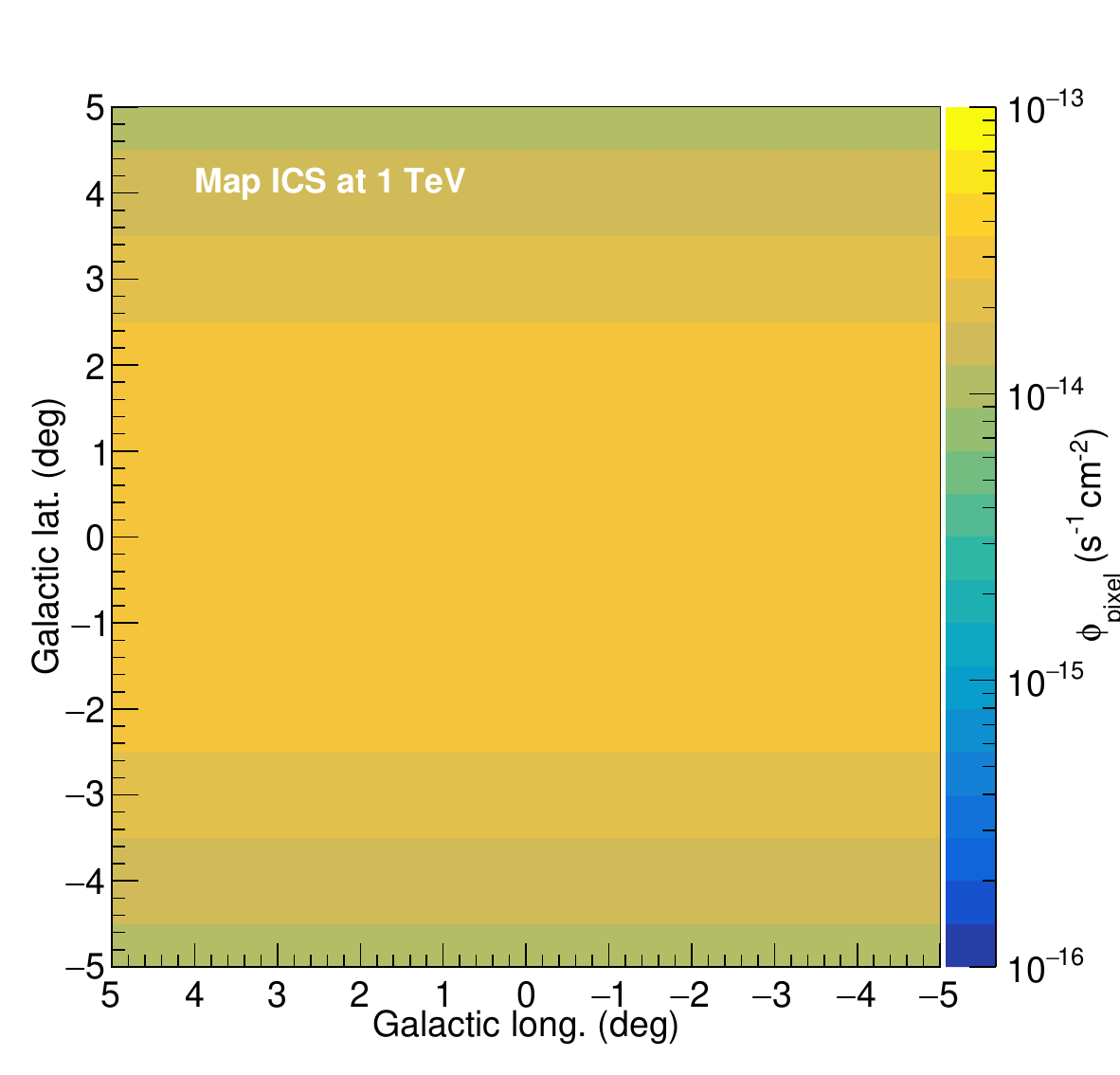}
\caption{Galactic diffuse emission maps in Galactic coordinates. The maps are expressed in terms of the flux $\Phi$ integrated in the energy bin centered at 1~TeV with width $\Delta\log_{10}(E/{\rm 1TeV})=0.2$, and in a  pixel of size $0.5^\circ\times0.5^\circ$.
The maps are given  for the GDE scenario 1 (top panels) and scenario 2 (bottom panels). 
The separate $\pi^0+$Brem (left panels) and ICS (right panels) components are shown.
 }
\label{fig:maps1TeV}
\end{center}
\end{figure*}

We employ also an alternative approach to estimating the diffuse emission relevant for this analysis, referred to as ``GDE scenario 2''.
For the second scenario we derive the spatial shape of $\pi^0$ emission using the measured distribution of interstellar dust throughout the Milky Way~\cite{Schlegel:1997yv}.
This dust is expected to approximately follow  the distribution of gas throughout the galaxy, which forms the targets for cosmic-ray protons to scatter off and form pions. 
To the extent this is true, and that further cosmic-ray protons are uniformly distributed, we can use this map as a proxy for the $\pi^0$ emission.
For the ICS, following Ref.~\cite{Su:2010qj}, we adopt a simple analytic shape as a function of Galactic longitude, $l$, and latitude, $b$, given by
\begin{equation}\begin{aligned}
{\rm ICS}(b,l) &\propto \exp \left[ - \frac{l^2}{2(30^{\circ})^2} \right] \\
&\times \left( \csc \left[ {\rm max}\left(2^{\circ},|b|\right)\right] - 1 \right)\,.
\end{aligned}\end{equation}
In the second line, the argument of the $\csc$ is not allowed to decrease for $|b| < 2^{\circ}$ in order to regulate the divergence at $b \to 0$.
To characterize the spectrum of the GDE components in this ``GDE scenario 2'', we perform a fit to the spectra given in the right panel of Fig. 1 in Ref.~\cite{TheFermi-LAT:2017vmf} (and plotted as empty triangles in the left panel of Fig.~\ref{fig:fluxes}, after translation to the ROI in question). We thus obtain a power-law spectrum for the gas-correlated emission and a power-law spectrum with an exponential cutoff for the ICS emission. We normalize the spatial maps in each energy bin so that their total fluxes within the ROI follow these spectra.

Reference~\cite{TheFermi-LAT:2017vmf} presented a detailed analysis of the emission near the Galactic center as observed by the {\it Fermi} satellite, and therefore can be used to estimate the diffuse emission relevant for CTA.
The data was fit to spatial templates in 27 logarithmically spaced energy bins between 100 MeV and 1 TeV.
The spatial templates included models for the diffuse emission (broken into $\pi^0$+
bremsstrahlung as well as ICS), isotropic emission, point sources, the Galactic center excess observed to peak around a few GeV, and a combined template for Loop I, Sun, Moon, and several other extended sources.
We use the fits performed in the smaller ROI considered in that work, which was $R < 10^{\circ}$ with $R$ the distance from the GC, as this is closer to our own ROI.
Finally, as the {\it Fermi} energy range does not extend as high as the expected CTA reach, we have extrapolated the relevant spectral dependences into the CTA energy range, using the parameters given in Tab.~\ref{tab:GDEmodel}.
The model is an extrapolation of the data points shown in the left panel of Fig.~\ref{fig:fluxes} (empty pink triangles).
The bottom panels of Fig.~\ref{fig:maps1TeV} show the spatial modeling of the GDE for this alternative scenario, in terms of integrated flux in the energy bin centered at 1~TeV and in each $0.5^\circ$-side-length square pixel.

\renewcommand{\arraystretch}{1.3}
\begin{table}[htbp]
\begin{center}
\begin{tabular}{c|c|c|c}
\hline
\hline
\multicolumn{4}{c}{Scenario 1}\\
\hline
Component & $\phi_0$ [TeV$^{-1}$cm$^{-2}$s$^{-1}$sr$^{-1}$] & $\Gamma$ & $E_{\rm cut}$ [TeV]\\
\hline
$\pi^0$ & $7\times10^{-10}$ & 2.48 & -\\
ICS & $3\times10^{-10}$ & 2.46 & 70\\
\hline
\hline
\multicolumn{4}{c}{ Scenario 2}\\
\hline
Component & $\phi_0$ [TeV$^{-1}$cm$^{-2}$s$^{-1}$sr$^{-1}$] & $\Gamma$ & $E_{\rm cut}$ [TeV]\\
\hline
$\pi^0$ & $8\times10^{-10}$ & 2.48 & -\\
ICS & $5\times10^{-10}$ & 2.40 & 100\\
\hline
\hline
\end{tabular}
\end{center}
\caption{Spectral parameters of the two models of the Galactic Diffuse Emission. The gas-correlated component is parametrized as a power-law and the ICS component as power-law with exponential cut-off.}
\label{tab:GDEmodel}
\end{table}

\subsection{{\it Fermi}-LAT High-Energy Sources}
\label{sec:3FHL}
The inner few degrees of the GC are populated by numerous high-energy gamma-ray sources that shine over the GDE~\cite{Fermi-LAT:2019yla}. Given that the highest energies are our focus, we select the point-like sources from the third Fermi-LAT high-energy source list~\cite{TheFermi-LAT:2017pvy} in Galactic longitude and latitude between $\pm$5$^\circ$, and for which the best-fit energy spectrum is a simple power law shape, {\it i.e.} with no indication of energy cut-off or break.

A disk of radius 0.25$^\circ$ centered at the nominal position of each selected source is used as a mask in order to avoid any modeling of the extrapolated spectral behaviour of these sources in the TeV energy range  that could be detected given the CTA sensitivity. Above 10 GeV, the {\it Fermi}-LAT PSF is about 0.1$^\circ$, therefore no significant leakage is expected outside the masked regions. The CTA PSF is even smaller. Masking all the selected 3FHL sources in the overall ROI degrades the overall solid angle of the search region by less than 1\%.
The dominant source of leakage is expected to be the uncertainty on the localization of the peak emission of faint and extended sources.
\begin{figure*}[htbp] 
\begin{center}
\includegraphics[width=0.45\textwidth]{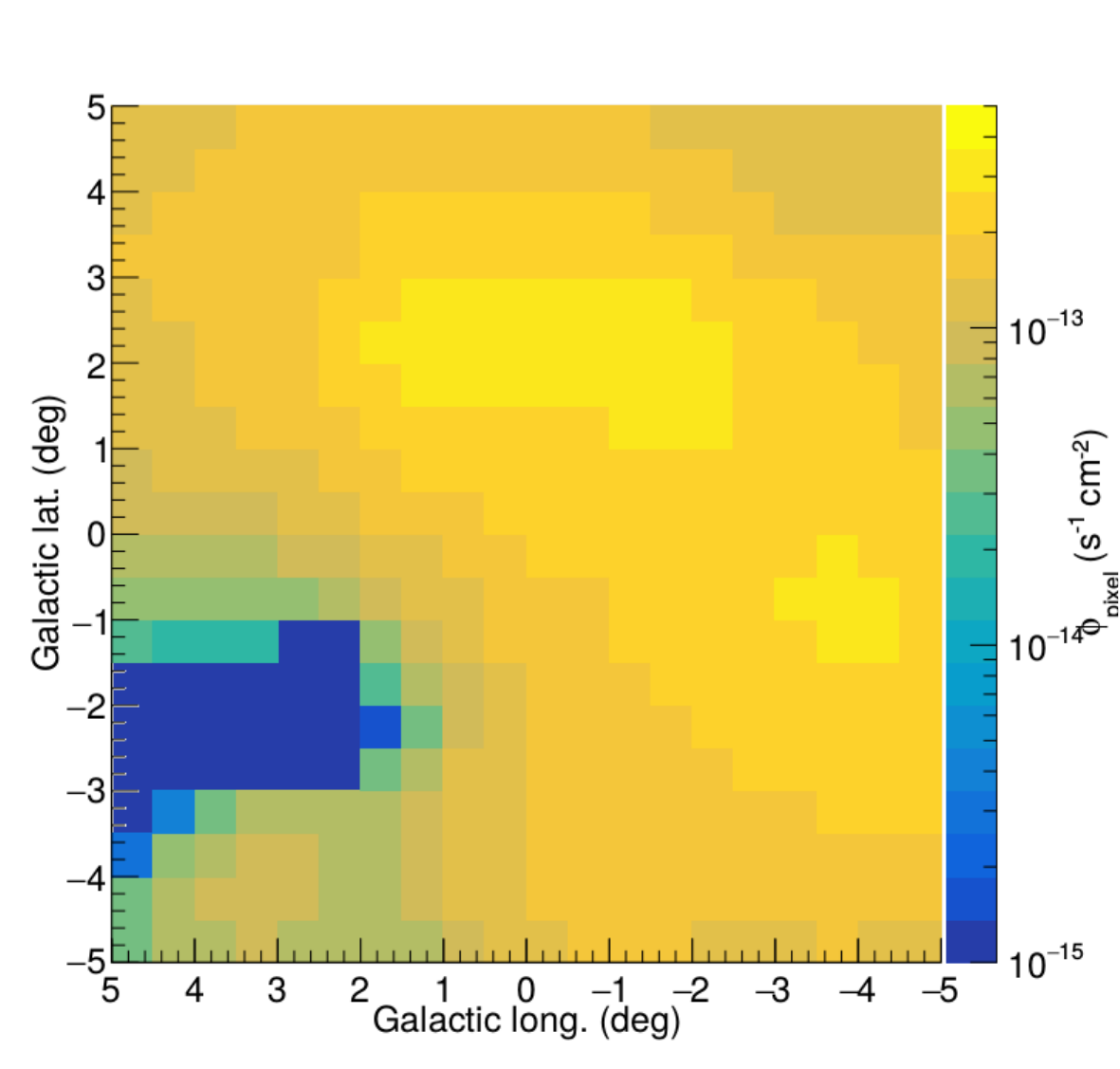}
\includegraphics[width=0.45\textwidth]{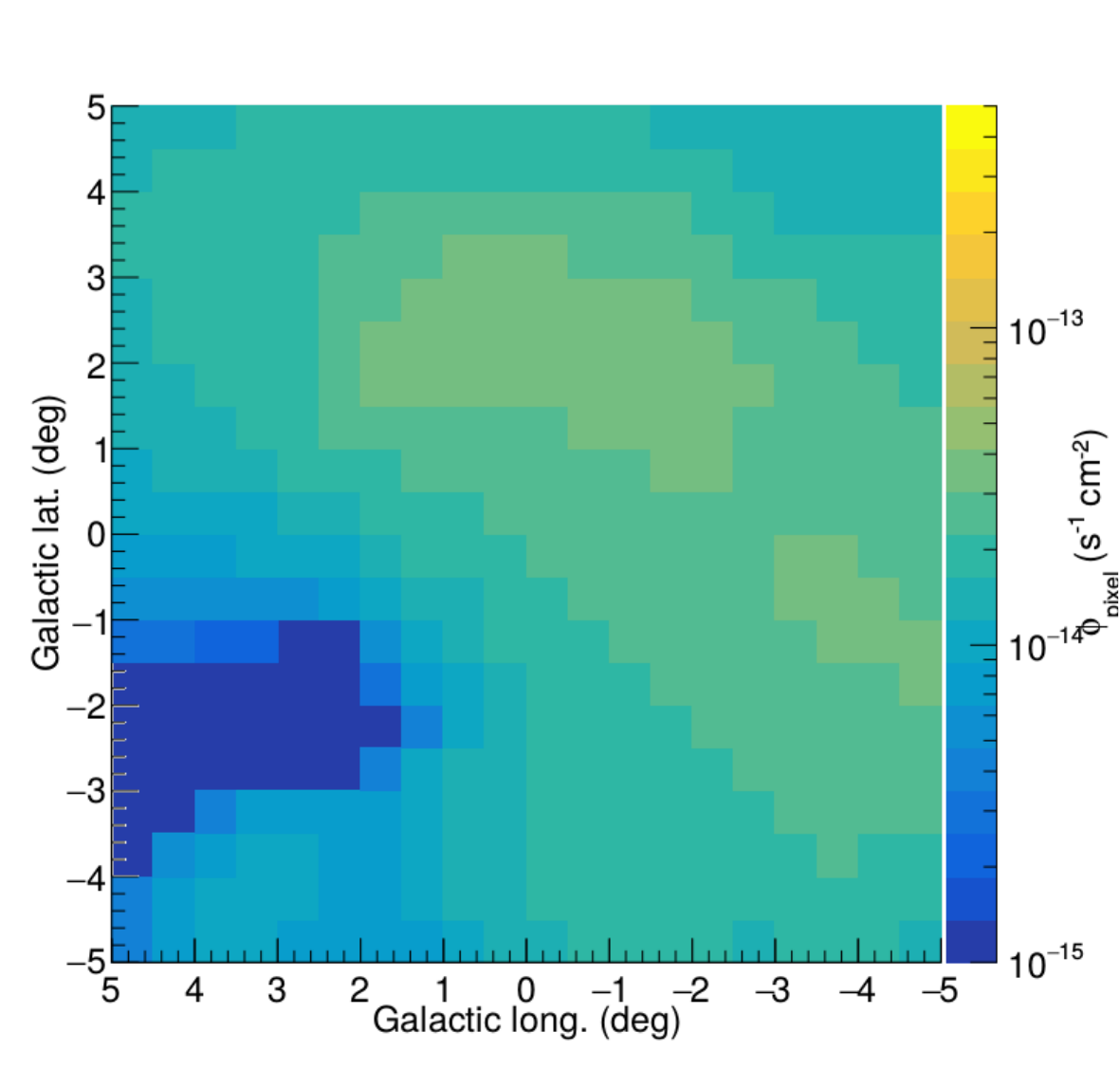}
\caption{Fermi Bubbles emission maps in Galactic coordinates for the conservative (left panel) and the optimistic (right panel) models. The maps are expressed in terms of flux integrated in a  pixel of size $0.5^\circ\times 0.5^\circ$ and  in the energy bin centered at 1 TeV with bin width $\Delta\log_{10}(E/{\rm 1TeV})=0.2$.}
\label{fig:FB1TeV}
\end{center}
\end{figure*}

\subsection{Fermi Bubbles at Low Galactic Latitudes}
\label{sec:FBs}
The Fermi Bubbles are large bipolar lobe structures, extending up to about 50$^\circ$ in Galactic latitudes above and below the Galactic plane, discovered in {\it Fermi}-LAT data~\cite{Su:2010qj}. At latitudes higher than 10$^\circ$, the Fermi Bubbles emission follows a $dN/dE \propto E^{-2}$ power law spectrum, with a significant spectral softening above 100 GeV. 
Recent analyses~\cite{Acero:2016qlg,TheFermi-LAT:2017vmf,Storm:2017arh,Herold:2019pei} indicate a brighter emission towards the Galactic plane ($|b|<10^\circ$) consistent with an $E^{-2}$ power-law spectrum that remains hard up to 1 TeV, {\it i.e.} no hints for an energy cut-off or break up to 1 TeV are detected.

In particular, Ref.~\cite{TheFermi-LAT:2017vmf} obtained a new spatial map for the low-latitude emission of the Fermi Bubbles by using an image reconstruction technique. This map was further improved in Ref.~\cite{Macias:2019omb} with the use of an inpainting method to correct for artifacts resulting from the point source mask applied in the analysis of~\cite{TheFermi-LAT:2017vmf}. A series of statistical tests were performed in Ref.~\cite{Macias:2019omb} to validate the improved low-latitude Fermi Bubbles map. In what follows we will assume the spatial template provided in Ref.~\cite{Macias:2019omb} as our template for the Fermi Bubbles.

The spectrum of the low-latitude component of the Fermi Bubbles is extracted from Ref.~\cite{TheFermi-LAT:2017vmf}. It is then extrapolated at energies above 1 TeV as an exponential cut-off power-law spectrum. Two models, ``FB min'' and ``FB max'', are considered with a spectral index $\Gamma = -1.9$~\cite{TheFermi-LAT:2017vmf}, respectively, and a normalization such that all the {\it Fermi} points fall between the two models.  The flux normalization and the energy cutoff of the optimistic and conservative models are  given in Tab.~\ref{tab:FBmodel}. The normalization of ``FB max'' (blue lines) is set in order to avoid overshooting the H.E.S.S. diffuse emission (red dots) as shown in the left panel of Fig.~\ref{fig:fluxes}, which shows the spectral behavior of the Fermi Bubbles as a function of the energy, together with the spectra of the GDE. 
Fig.~\ref{fig:FB1TeV} shows the spatial behavior of the Fermi Bubbles low-latitude emission for the ``FB min" (left panel) and ``FB max" (right panel) models. The flux is integrated in the 1~TeV energy bin and in each $0.5^\circ$-side-length pixel.  
\begin{table}[htbp]
\begin{center}
\begin{tabular}{c|c|c|c}
\hline
\hline
Model & $\phi_0$ [TeV$^{-1}$cm$^{-2}$s$^{-1}$sr$^{-1}$] & $\Gamma$ & $E_{\rm cut}$ [TeV]\\
\hline
FB max & $1\times10^{-8}$ & 1.9 & 20\\
FB min & $0.5\times10^{-8}$ & 1.9 & 1\\
\hline
\hline
\end{tabular}
\end{center}
\caption{Spectral parameters for the two parametrizations of the low-latitude component of the Fermi Bubbles emission, modeled as a power-laws with an exponential cut-off in energy. The parameters are given for the optimistic ``FB max" and the conservative ``FB min" models, respectively.}
\label{tab:FBmodel}
\end{table}

\section{Sensitivity}
\label{sec:sensitivity}
\subsection{Instrument Response Functions}
The Southern site of the CTA observatory is best-suited to observe the GC region under the most favorable observation conditions.
The IRFs of CTA used in this work are obtained from publicly available Monte Carlo simulations with an array composed of 4 large-size telescopes,  24 medium-size telescopes and 70 small-size telescopes, optimized for 50 hour observation time~\cite{CTAperformances}. 
The energy-dependent acceptance and residual background rate, together with the angular and energy resolution, are computed in the energy range from 10 GeV -- 100 TeV, for an observation zenith angle of 20$^\circ$. We note that an energy resolution as low as 5\% can be achieved at TeV energies.  
  
The version prod3b-v1 of the IRFs are chosen for {\it on-axis} observations, {\it i.e.} for a source localized close to the center of the field of view. As shown in Ref.~\cite{CTAperformances}, the sensitivity deteriorates by less than a factor of two for sources up to 2.5$^\circ$ from the center of the field of view,  for energies higher than 1 TeV. In what follows we assume a homogeneous sky exposure of 500 hours in the overall search region, which can be obtained provided that an ambitious observation program of the inner Galactic halo is carried out with an optimized observation strategy~\cite{Moulin:2019oyc}.

\subsection{Expected Signal and Background Events}\label{sec:sigback}
In order to obtain an estimate of the expected signal $s_{\gamma, ijk}$ in the $i^{\rm th}$ energy, $j^{\rm th}$ Galactic longitude and $k^{\rm th}$ Galactic latitude   bins, the differential gamma-ray flux ${\rm d}\Phi_{\gamma, jk}^{\rm S}/{\rm d} E_\gamma$ in the bin $ijk$, integrated over the dimensions of the ROI, is convolved with the CTA gamma-ray acceptance $A_{\rm eff}^\gamma$ and energy resolution at energy $E_\gamma'$: 
\begin{equation}
\frac{d\Gamma_{\gamma, jk}^{\rm S}}{d E_\gamma}(E_\gamma)= \int_{-\infty}^{+\infty} d E_\gamma'  \frac{d\Phi_{\gamma, jk}^{\rm S}}{d E_{\gamma}}(E_\gamma') \, A^{\gamma}_{\rm eff}(E_\gamma') \, G(E_\gamma,E_\gamma').
\label{eq:signalrate}
\end{equation}
The energy resolution is modeled as a Gaussian $G(E_\gamma,E'_\gamma)$ with width (68\% containment radius) of 20\% at 50 GeV, down to better than 5\% in the TeV energy range~\cite{CTAperformances}. The signal count number $s_{\gamma, ijk}$ in the bin $ijk$
is obtained from the gamma-ray differential rate $d\Gamma_{\gamma, jk}^{\rm S}/d E_\gamma$  integrated over the energy bin $\Delta E$ and multiplied by the observation time $T_{\rm obs}$. Explicitly, \begin{equation}
s_{\gamma, ijk}=T_{\rm obs} \int_{\Delta E_i}  dE_\gamma \, \frac{d\Gamma_{\gamma, jk}^{\rm S}}{d E_\gamma} \, .
\label{eq:signacount}
\end{equation}
The modeled background $b_{\gamma, ijk}$ in the bin $ijk$ is obtained:
\begin{equation}
b_{\gamma, ijk}=T_{\rm obs} \int_{\Delta E_i}  dE_\gamma \, \frac{d\Gamma_{\gamma, jk}^{\rm B}}{d E_\gamma} \, .
\label{eq:bckcount}
\end{equation}
The CR background flux $d\Phi_{\gamma, jk}^{\rm CR}/d E_{\gamma}d \Omega$ is multiplied by the CR acceptance of CTA, $A_{\rm eff}^{\rm CR}$, while the standard background flux due to astrophysical gamma-ray sources $d \Phi_{\gamma, jk}^{\rm Std}/d E_{\gamma}d \Omega$ is multiplied by $A_{\rm eff}^\gamma$:
\begin{align}
\frac{d\Gamma_{\gamma, jk}^{\rm B}}{d E_\gamma}&(E_\gamma)=
 \int_{-\infty}^{+\infty} d E_\gamma' \int_{\Delta\Omega_{jk}} d\Omega\, \Bigg[\frac{d \Phi_{\gamma, jk}^{\rm CR}}{d E_{\gamma}d \Omega}(E_\gamma') \, A^{\rm CR}_{\rm eff}(E_\gamma') \nonumber \\ 
 & +  \frac{d \Phi_{\gamma, jk}^{\rm Std}}{d E_{\gamma}d \Omega}(E_\gamma', \Delta\Omega)A^{\gamma}_{\rm eff}(E_\gamma')\Bigg] \, G(E_\gamma,E_\gamma').
\label{eq:bckrate}
\end{align}
The gamma-ray backgrounds due to the point-like sources $d\Phi_{\gamma, jk}^{\rm PL}/d E_{\gamma}d\Omega$,
the GDE  $d\Phi_{\gamma, jk}^{\rm GDE}/d E_{\gamma}d\Omega$, and the FB $d\Phi_{\gamma, jk}^{\rm FB}/d E_{\gamma}d \Omega$ are included in $b_{\gamma, ijk}$:

\begin{align}
\frac{d \Phi_{\gamma, jk}^{\rm Std}}{d E_{\gamma}d \Omega}(E_\gamma', \Delta\Omega) =& \frac{d \Phi_{\gamma, jk}^{\rm PL}}{d E_{\gamma}d \Omega}(E_\gamma', \Delta\Omega) + \frac{d \Phi_{\gamma, jk}^{\rm GDE}}{d E_{\gamma}d \Omega}(E_\gamma', \Delta\Omega) \nonumber\\ 
&+ \frac{d \Phi_{\gamma, jk}^{\rm FB}}{d E_{\gamma}d \Omega}(E_\gamma', \Delta\Omega).
\label{eq:bkg}
\end{align}

In actual observations with IACTs, typically a fit including both signal and background components is performed in a signal region (known as the ON region), and simultaneously the background is constrained by observations of a corresponding control region (OFF region) where no signal is expected. This approach has the advantage of factoring out systematic effects associated with the instrument, which should affect the background in the OFF and ON regions similarly. However, it depends on being able to select an appropriate OFF region with similar background to the ON region, which may be challenging if there is a significant detection of the spatially non-uniform GDE in CTA observations. In this work, we do not aim to address the question of the optimal observation strategy, the choice of telescope pointing positions, and the definition of the OFF regions; instead we use a simplified approach where we simply model the expected background in the ON region. This is equivalent to positing an OFF region that perfectly constrains the background model. In a real analysis, the noise in the measurement of the background from the OFF region would be expected to degrade the signal sensitivity somewhat, compared to our simplified analysis; alternatively, using our background-modeling approach with real data would require consideration of possible systematic errors from mis-modeling of either the instrument response or the GDE itself \cite{Silverwood:2014yza}. In this sense, our results represent a best-case scenario for sensitivity to a Wino or Higgsino signal.

Fig.~\ref{fig:fluxes} shows the expected count rate as a function of the energy in a central ROI for the residual background, the two GDE models, the ``FB max'' model, and expected signals from Wino and Higgsino DM, assuming the corresponding thermal mass and predicted annihilation cross section. 
\begin{figure*}[htbp] 
\begin{center}
\includegraphics[width=0.45\textwidth]{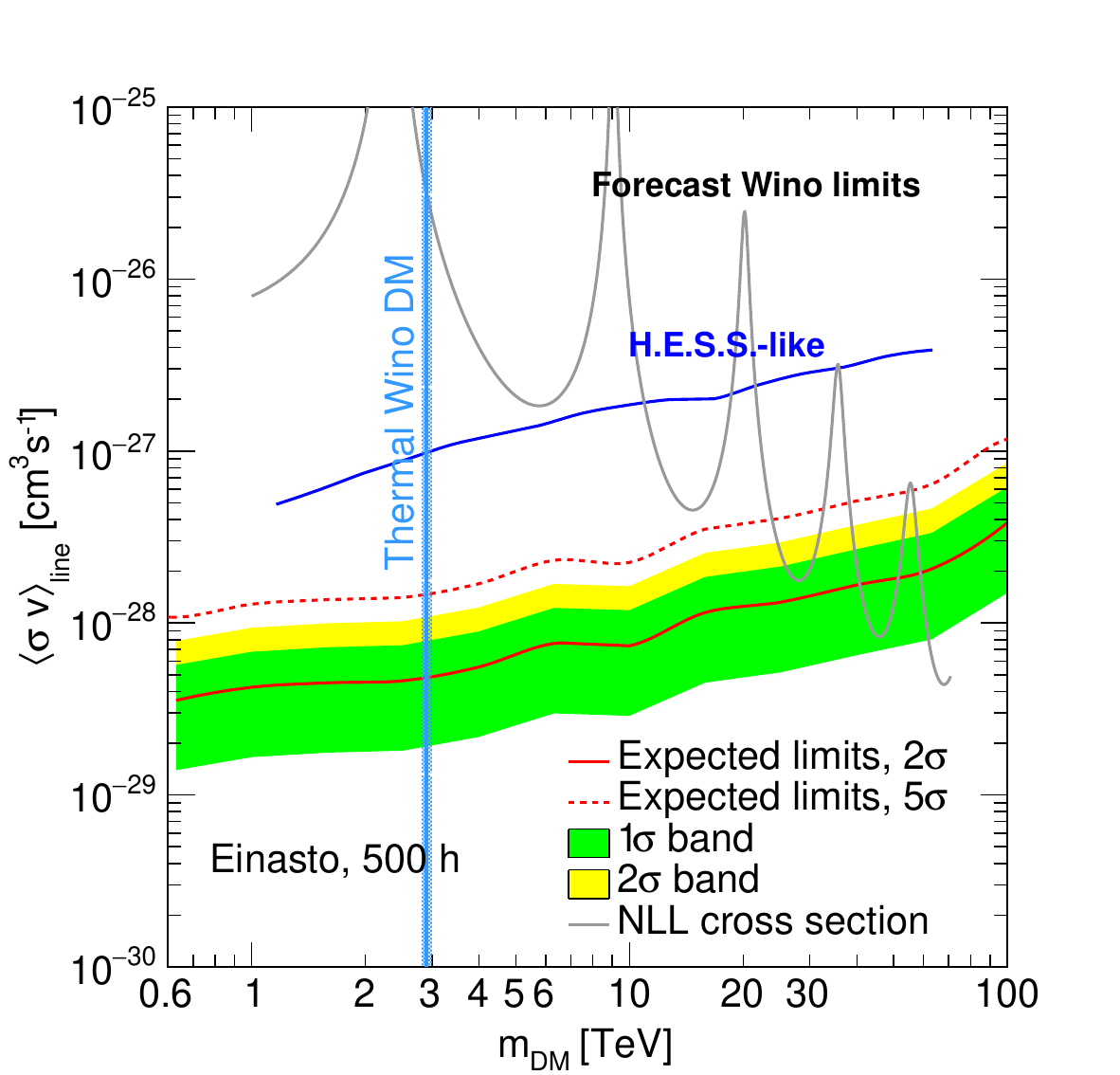}
\includegraphics[width=0.45\textwidth]{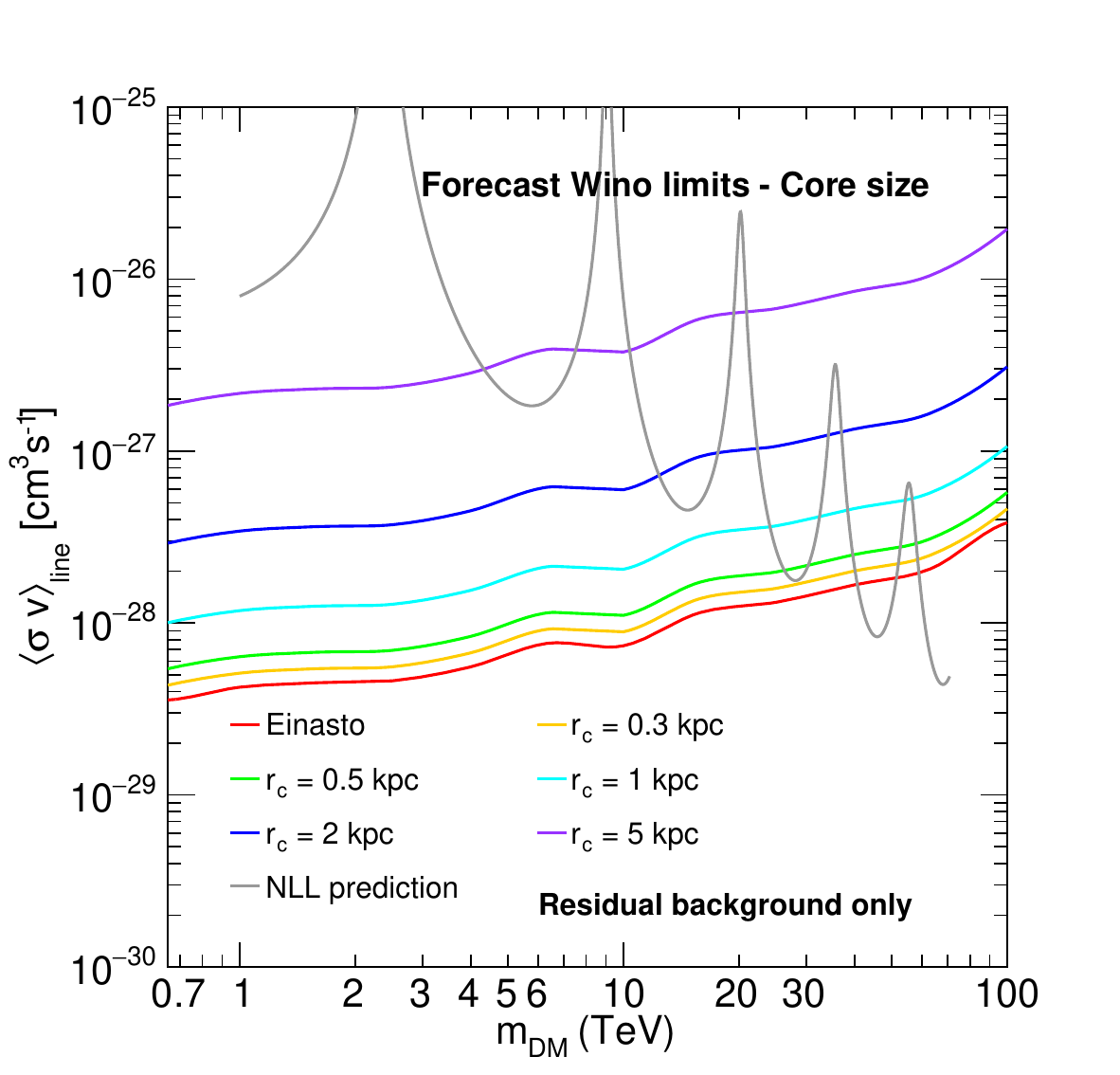}
	\caption{Expected upper limits at 95\% C.L on the Wino annihilation cross section as a function of its mass for 500 h of CTA observations towards the GC. The predicted NLL cross section is shown (solid gray line) and the thermal Wino DM mass is marked (cyan solid line and bands). The only  background considered here is the residual background. The full Wino spectrum is included in the expected signal. {\it Left panel:} Mean expected upper limits at 2$\sigma$ (red solid line) for an Einasto profile are shown together with the $1\sigma$ (green band) and $2\sigma$ (yellow band) containment bands. 
Mean expected upper limits at 5$\sigma$ (red dashed line) are also shown. The H.E.S.S.-like 2$ \sigma$ sensitivity extracted from Ref.~\cite{Rinchiuso:2018ajn} is shown as a blue solid line. {\it Right panel:} The expected limits are shown for cored DM profiles of size from 300~pc to 5~kpc.}
\label{fig:WinoBandsCores}
\end{center}
\end{figure*}

\subsection{Sensitivity Computation}
A 3D likelihood-ratio test statistic technique is used to compute the CTA sensitivity to $\langle\sigma v\rangle_{\rm line}$ in the Higgsino and Wino DM models. A standard likelihood function for a counting experiment is used. The likelihood in the bin $i,j,k$ is given by the Poisson distribution:\footnote{If an ON-OFF approach is used, a second Poisson factor for the OFF region is added, multiplying that for the signal ON region.}
\begin{equation}\begin{aligned}
&\mathcal{L}_{ ijk}(s_{\gamma, \rm ijk}+b_{\gamma, \rm ijk}, m_{\gamma, \rm ijk}) \\
=\,&\text{Pois}(s_{\gamma, \rm ijk}+b_{\gamma, \rm ijk}, m_{\gamma, \rm ijk}).
\label{eq:lik}
\end{aligned}\end{equation}
where $\text{Pois}(\lambda,p) = e^{-\lambda}\,\lambda^p/p!$, $s_{\gamma, \rm ijk}$ and $b_{\gamma, \rm ijk}$ describe the expected photon number from the signal and background models respectively, and $m_{\gamma, \rm ijk}$ represents the observed photon number in the relevant bin (or an appropriate proxy; see the discussion of the Asimov method we employ below).
For the sensitivity studies in the present work, we keep $b_{\gamma, \rm ijk}$ fixed in the form of the model derived from
Eqs.~(\ref{eq:bckrate}-\ref{eq:bkg}),
so that our background model contains no free parameters. As such, once the DM mass and model (e.g. whether it is a Wino or Higgsino) is specified, the only free parameter in the signal model and likelihood is an overall signal normalization factor controlled by $\langle \sigma v \rangle_\text{line}$.

The likelihood function is binned in energy (indexed by $i$), Galactic longitude (indexed by $j$) and Galactic latitude (indexed by $k$). The total likelihood is the product of $\mathcal{L}_{ ijk}$ over the 20 energy bins and 400 spatial bins. 
In our case the background $b_{\gamma, \rm ijk}$ is modeled rather than being measured in an OFF region, as explained in Sec.~\ref{sec:sigback}, and as mentioned above the background model contains no free parameters (we do not allow its normalization, for example, to vary).\footnote{The future  telescope pointing strategy of CTA that will be implemented to survey the GC region will define optimized pointing positions of the telescopes to most efficiently survey the GC region, together with the OFF regions where the background will be measured for each observation. This discussion is beyond the scope of this work.}
~The sensitivity is expressed here as the expected limit obtained under the assumption that $m_{\gamma, \rm ijk}$ contains no DM signal. 
Values of $\langle\sigma v\rangle_{\rm line}$ are tested through the likelihood ratio test statistic profile defined as:
\begin{equation}
\Lambda_{ijk} = \frac{\mathcal{L}_{ ijk}(s_{\gamma, \rm ijk}+b_{\gamma, \rm ijk}, m_{\gamma, \rm ijk})}{\mathcal{L}_{ ijk}(\hat{s}_{\gamma, \rm ijk}+b_{\gamma, \rm ijk}, m_{\gamma, \rm ijk})}.
\label{eq:teststatistics}
\end{equation}
In the ratio, only the amplitude of $s_{\gamma, \rm ijk}$ is a free parameter, and therefore this quantity is solely a function of the cross section $\langle \sigma v \rangle_{\rm line}$. In the denominator we fix the signal flux normalization to the value which maximises the likelihood, denoted by $\hat{s}_{\gamma, \rm ijk}$. Using Eq.~(\ref{eq:teststatistics}), we can then define a test statistic for setting upper limits as
\begin{equation}
q(\langle \sigma v \rangle)
= \left\{\begin{array}{lc} -\sum_{ijk}2\ln\Lambda_{ijk} & \langle \sigma v \rangle \geq \widehat{\langle \sigma v \rangle}\,, \\
0 & \langle \sigma v \rangle < \widehat{\langle \sigma v \rangle}\,,
\end{array}\right.
\end{equation}
where the cross section is again $\langle \sigma v \rangle_{\rm line}$, and here $\widehat{\langle \sigma v \rangle}$ corresponds to the value of the cross section where the best fit signal is achieved, in detail the value that determined $\hat{s}_{\gamma, \rm ijk}$ as in the denominator of Eq.~(\ref{eq:teststatistics}).
As the cross section is increased, eventually the signal strength will become incompatible with the data and $q$ will begin to increase.
The value of $\langle\sigma v\rangle_{\rm line}$ excluded at 95\% confidence level corresponds to $q \approx 2.71$, when computing one-sided upper limits, assuming that the test statistic behaves as $\chi^2$ distribution, as expected in the high statistic limit, with one degree of freedom.
Note that this prescription uses Wilks' theorem, and as such requires that we allow $\langle\sigma v\rangle_{\rm line}$ to float negative, as if the background fluctuates below its mean, the best fit signal point can be negative.
This choice implies that for a significant enough downward background fluctuation, we could potentially set a negative limit on $\langle\sigma v\rangle_{\rm line}$, thereby excluding all positive values.
In order to avoid this possibility, we implement power-constrained limits as described in Ref.~\cite{Cowan:2011an}, where the actual limit is not allowed to go below the lower 1$\sigma$ expected limit.
Accordingly, in Figs.~\ref{fig:WinoBandsCores} and \ref{fig:Higgsinocores}, we only show the lower 1$\sigma$ expected limit, as the actual limit, by construction, cannot go below this.
We also compute the 5$\sigma$ mean expected upper limit on $\langle\sigma v\rangle_{\rm line}$, which corresponds to 
q $\approx$ 23.7.} 

The above prescription outlines how to determine the limit for a given dataset $m_{\gamma, \rm ijk}$, which could be either obtained from real observations or via Monte Carlo simulations. 

\begin{figure}[!t] 
\begin{center}
\includegraphics[width=0.45\textwidth]{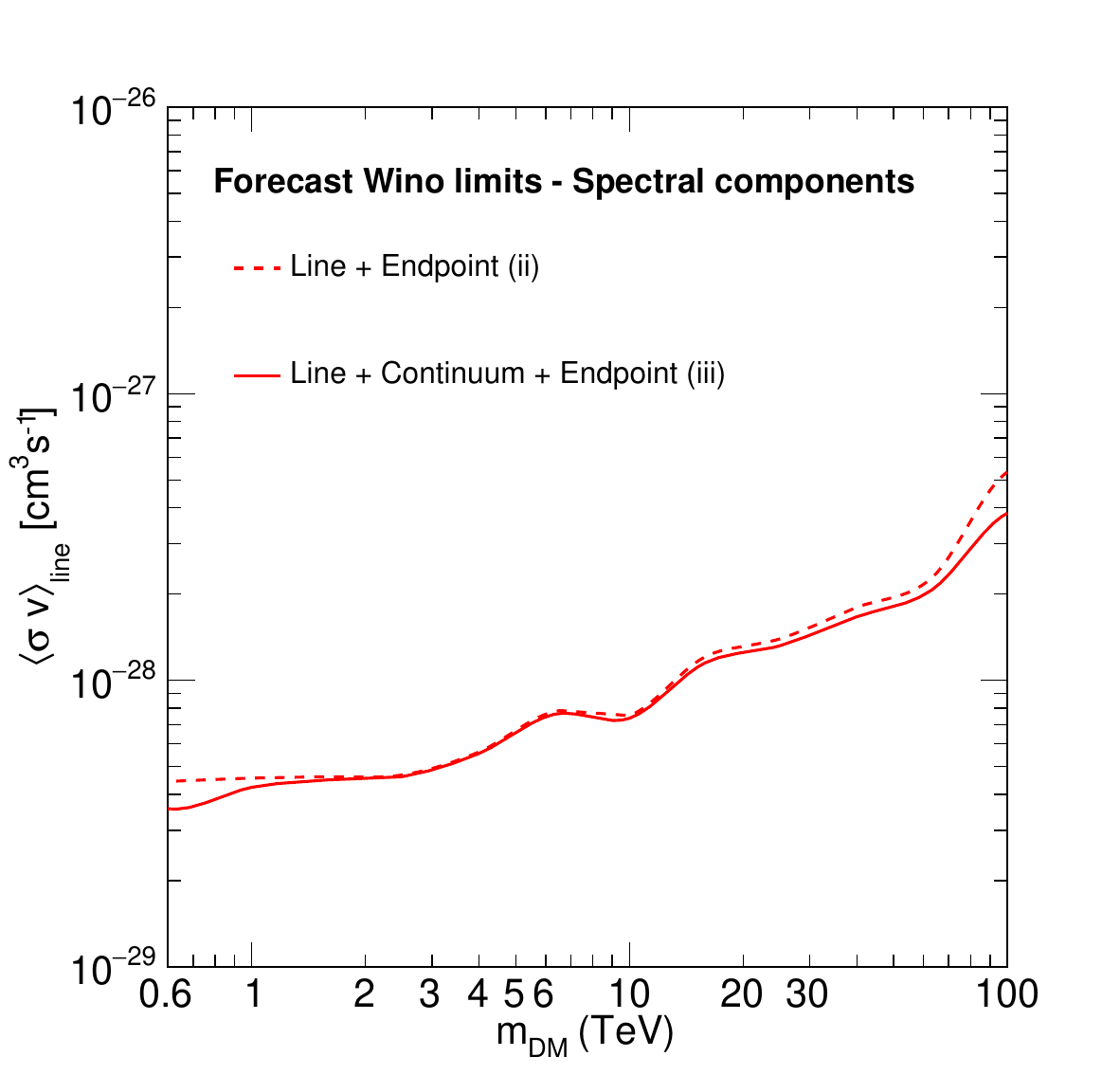}
\caption{Impact on the 95\% C.L. mean expected  upper limits, on the Wino annihilation cross section, of the continuum and endpoint contributions to the Wino spectrum. The limits for the full Wino Spectrum (red solid line) are compared to those that do not include the continuum contribution (red dashed line). The major improvement on the sensitivity to the Wino with respect to a simple line comes from the endpoint gamma-rays. The effect of the continuum on the limits is much smaller than the endpoint contribution, to a greater degree than observed with H.E.S.S., due to the improved CTA energy resolution.}
\label{fig:endpointcontribution}
\end{center}
\end{figure}
Before CTA's first light, we can estimate the expected sensitivity by generating a large number of Monte Carlo datasets and determining the mean expected limit and associated containment bands.
An alternative to this approach, which we will use in this work, is to instead determine all of these quantities using the Asimov formalism of Ref.~\cite{2011EPJC711554C}.
Under the Asimov approach, instead of taking many realizations of the model, calculating the limit each time, and then determining the mean of those values, we instead take the mean dataset, which is exactly given by the model.
The model, when used as the dataset, is then referred to as the Asimov dataset.
Of course, as the model is not strictly an integer, this requires analytically continuing the Poisson distribution to non-integer values, which can be accomplished using the $\Gamma$ function.
The Asimov approach can also be used to determine the confidence intervals.
In detail, to determine the $N$-sigma containment band, instead of evaluating $q =2.71$, we calculate
\begin{equation}
q = \left( \Phi^{-1}(0.95) \pm N \right)^2\,.
\label{eq:tstot}
\end{equation}
Here $\Phi$ is the cumulative distribution function for the standard normal, which has $\mu=0$ and $\sigma=1$.
Accordingly $\Phi^{-1}(0.95) \approx 1.64$, so that the above result contains the mean limit as a special case at $N=0$.

In the idealized scenario we consider here of data drawn from a background model known exactly, the above procedure for calculating limits is sufficient.
We emphasize, however, that when considering the actual CTA data, our models will be inevitably imperfect.
One consequence of this is that the coverage of our limits, and the validity of discovery thresholds can deviate from the simple asymptotic estimates used above, and may need to be validated and potentially tuned using datasets that contain an injected signal.

\section{Results and Prospects}
\label{sec:results}
\subsection{Sensitivity to Wino DM and impact of the endpoint contribution}
The CTA sensitivity forecast for Wino DM, expressed as the mean expected upper limit at 95\% C.L. on $\langle\sigma v\rangle_{\rm line}$ as a function of the Wino mass, is shown in the left panel of Fig.~\ref{fig:WinoBandsCores}, together with the expected containment bands obtained from the Asimov dataset. This forecast assumes 500 h of homogeneous exposure within the overall $10^\circ$-side ROI centered on the GC, and an Einasto DM density profile. The Wino spectrum includes the line and the continuum component together with the endpoint contribution.
The mean expected limits (red solid line) reach cross sections of about $4\times10^{-29}$~cm$^3$~s$^{-1}$ at 1~TeV. The containment bands at $1\sigma$ (green band) and $2\sigma$ (yellow band) are drawn in green and yellow, respectively, together with the theoretical cross section (gray solid line). 
All masses where the CTA limits lie below the theoretical cross section are forecast to be excluded. The mass $m_{\DM}=2.9$~TeV predicted for a Wino thermally-produced in the early universe can be excluded by CTA; this is not surprising, as it has already been shown that this mass and cross section can be strongly constrained by H.E.S.S.-like observations~\cite{Rinchiuso:2018ajn}. For these assumptions, the Wino cross section can be probed up to masses of $\sim$40~TeV outside resonances.

Less stringent constraints are obtained if a cored DM distribution is assumed in the GC, as shown in the right panel of Fig.~\ref{fig:WinoBandsCores}. The sensitivity to $\langle\sigma v\rangle_{\rm line}$ is computed for DM density cores of radius $r_{\rm c}=$~0.3, 0.5, 1, 2 and 5~kpc. 
The degradation of the limits is only a factor 20\% for a 300~pc ($\simeq$2.0$^\circ$) core, but increases up to a factor of about 50 in the extreme case of a core with $r_{\rm c}$ = 5 kpc ($\simeq$30.5$^\circ$). The Wino thermal mass is expected to be excluded even for the largest cores. Masses below 4~TeV, 9~TeV and 20~TeV at the positions of Sommerfeld resonances are excluded for cores as large as 5~kpc. Only masses above a few ten TeV are out of reach for kpc-sized cores outside resonances. The 5$\sigma$ sensitivity is plotted in the left panel of Fig.~\ref{fig:WinoBandsCores}.

Alternatively, if we assume our fiducial Einasto profile, we can interpret the results of Fig.~\ref{fig:WinoBandsCores} as constraining the fraction of DM that can consist of 2.9~TeV Winos.
CTA will exclude the thermal prediction by a factor of $\sim$620, so a null detection would allow the Wino to amount to no more than 4\% of DM (recall that for annihilation the signal flux scales quadratically with the DM fraction).
For masses below the thermal value of 2.9~TeV, the Wino is naturally only a subset of the full DM density.
CTA would be able to test even these scenarios: a 1~TeV Wino would contribute $\sim$17\% of DM~\cite{Cirelli:2007xd}, whereas the cross-section limit in Fig.~\ref{fig:WinoBandsCores} requires such DM to constitute a fraction no larger than 7\%.

Fig.~\ref{fig:endpointcontribution} shows the improved sensitivity obtained by using the full Wino annihilation spectrum, including the continuum and endpoint contributions, in addition to the monoenergetic gamma line. The endpoint component improves the constraints by a factor of 1.5 at 1~TeV, 3 at 10~TeV and 7 at 50~TeV. Inclusion of the continuum improves the limits by 1-5\% over the whole DM mass range. 
The improvement due to the continuum is less pronounced than expected for H.E.S.S. observations~\cite{Rinchiuso:2018ajn}, whereas the endpoint contribution is more significant. This difference is due to the improved energy resolution of CTA compared to H.E.S.S., which is up to a factor of two better.
The line and endpoint contributions to the overall spectrum strongly dominate the expected limits. 

\subsection{Sensitivity to the Higgsino}
\begin{figure*}[!htbp] 
\begin{center}
\includegraphics[width=0.45\textwidth]{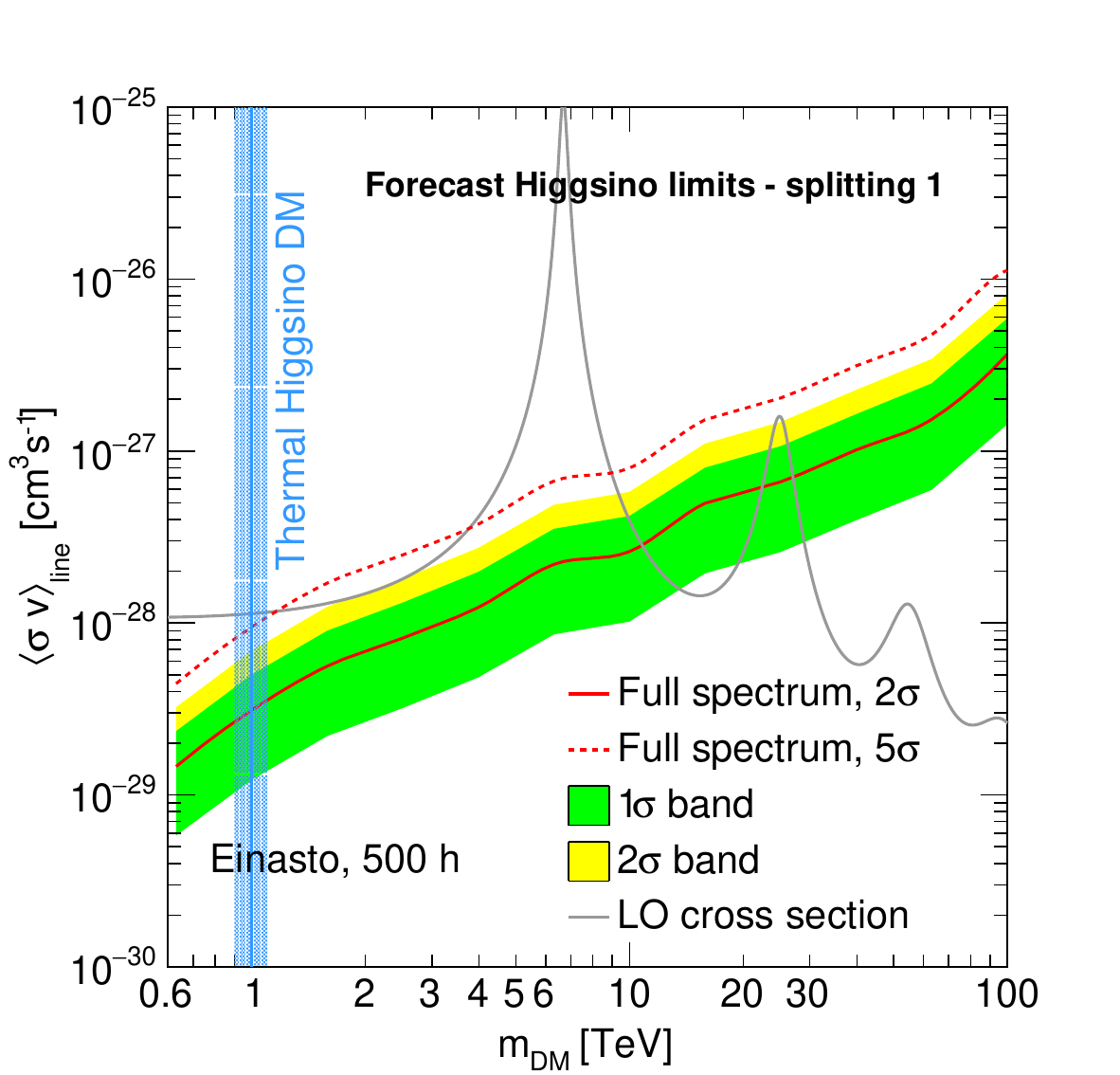}
\includegraphics[width=0.45\textwidth]{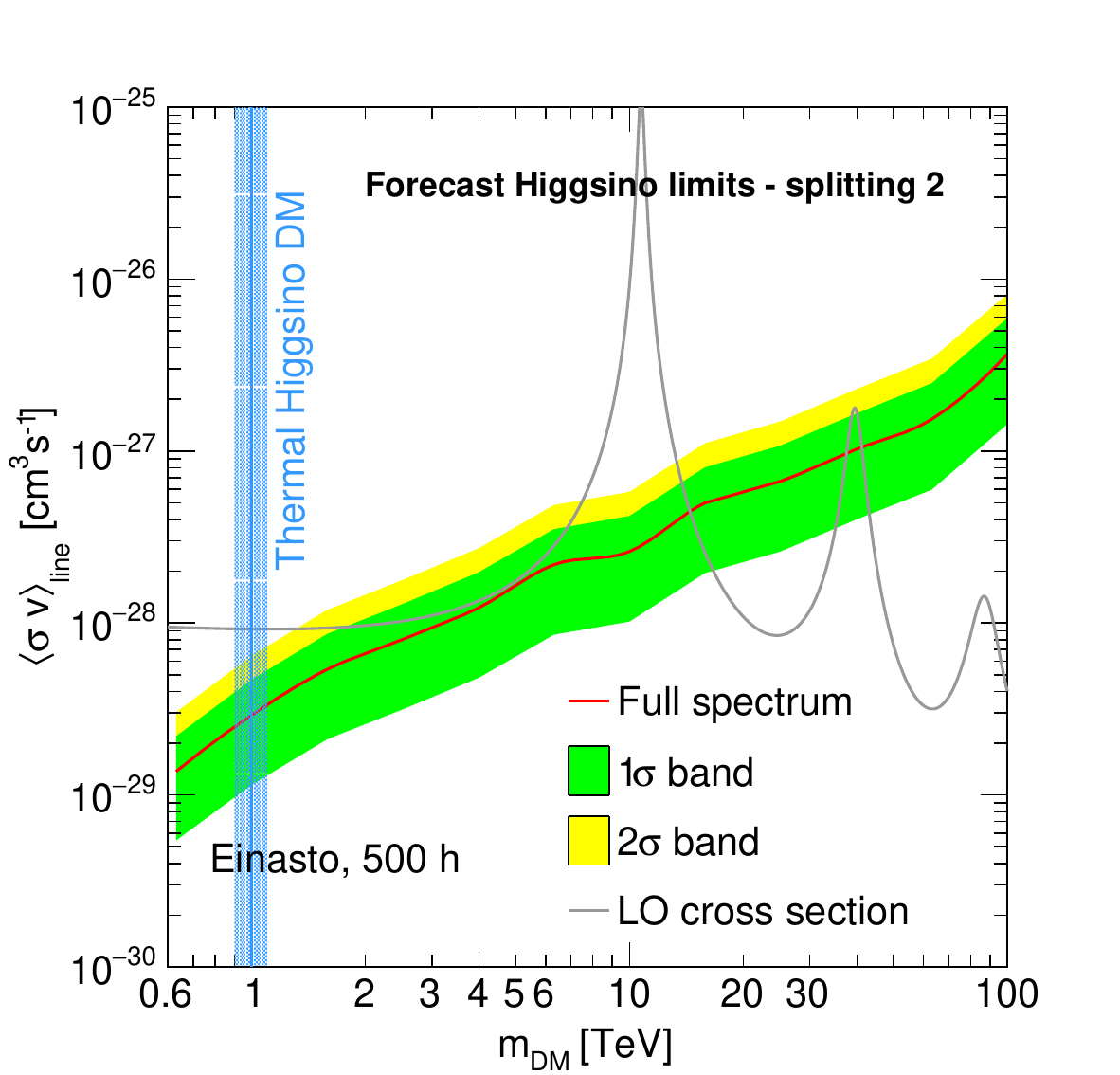}
\includegraphics[width=0.45\textwidth]{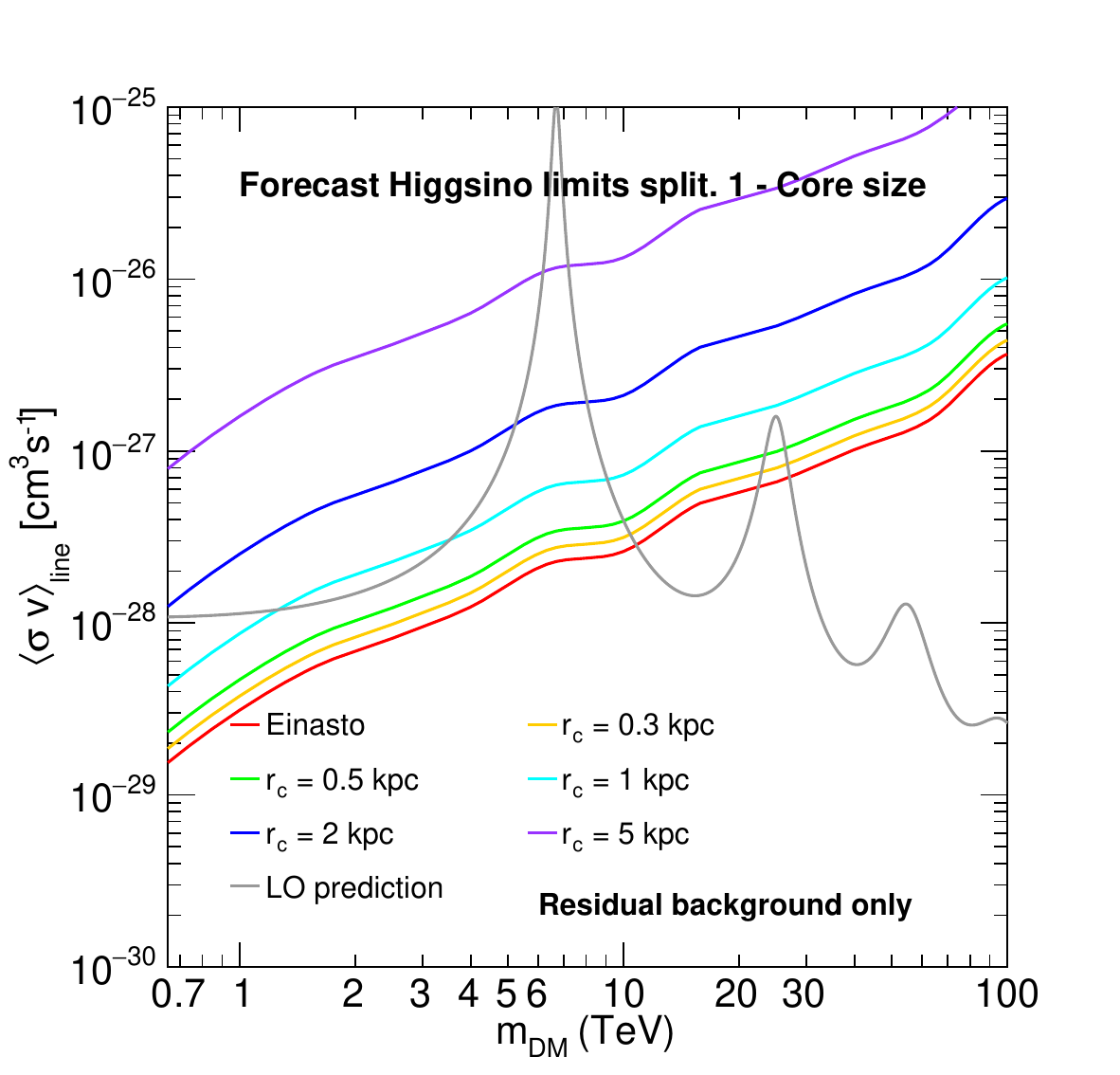}
\includegraphics[width=0.45\textwidth]{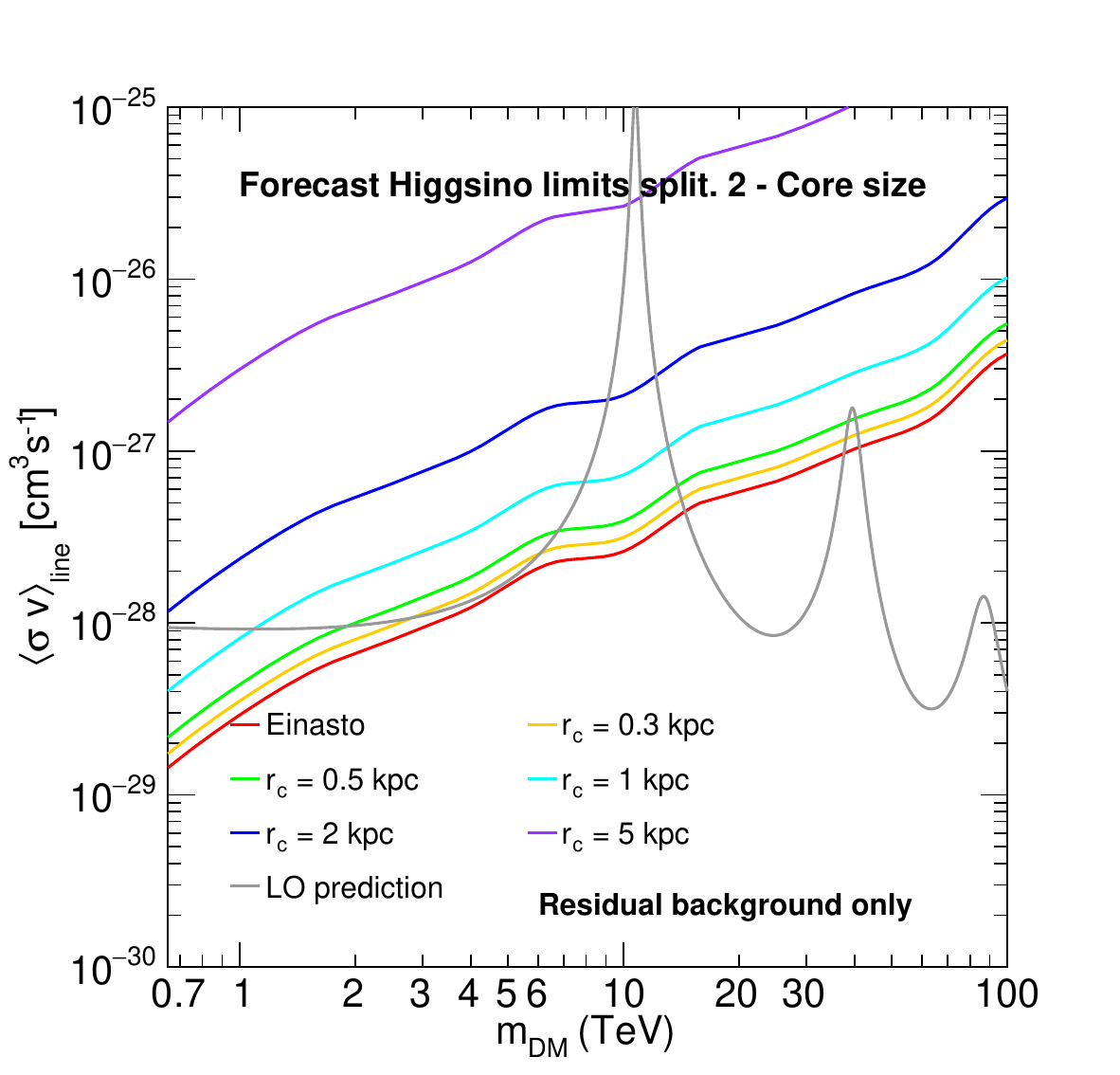}
\caption{
95\% C.L. expected upper limits on the line Higgsino annihilation cross section as a function of its mass for the Einasto profile (red solid line) and cores of size from 300~pc to 5~kpc. The theoretical cross section is printed in gray. {\it Top left panel:} Limits computed assuming mass splittings $\delta m_N=200$~keV and $\delta m_+=350$~MeV. The mean expected limits are shown at 2$\sigma$ (red solid line) and 5$\sigma$ (red dashed line), respectively. 
{\it Top right panel}: Limits computed assuming mass splittings $\delta m_N=2$~GeV and $\delta m_+=480$~MeV.  {\it Bottom panels:} 95\% C.L. expected mean upper limits for CTA on the Higgsino annihilation cross section as a function of its mass, for an Einasto DM profile and 500~hour homogeneous exposure in a $10^\circ$-side squared region centered at the GC region. The expected limits (red solid line) are shown together with the $1\sigma$ (green band) and $2\sigma$ (yellow band) containment band obtained from the Asimov dataset. Only the residual background is considered here. The predicted LO cross section is shown (solid gray line) and the thermal Higgsino DM mass is marked (cyan solid line and bands). The sensitivity is computed for the mass splittings $\delta m_N=200$~keV and $\delta m_+=350$~MeV (bottom left panel) and $\delta m_N=2$~GeV and $\delta m_+=480$~MeV (bottom right panel). 
 The line-only constraints are shown as red dotted lines.}
\label{fig:Higgsinocores}
\end{center}
\end{figure*}

The CTA sensitivity forecast for Higgsino DM, expressed as the expected 95\% C.L. upper limit on $\langle\sigma v\rangle_{\rm line}$ as a function of the Higgsino mass, is shown in Fig.~\ref{fig:Higgsinocores}, together with the expected containment bands obtained from the Asimov dataset. The theoretical cross section is overlaid in gray. Two cases are considered for different splittings: ``splitting 1'' (left panel) refers to $\delta m_N=200$~keV and $\delta m_+=350$~MeV and ``splitting 2'' (right panel) to $\delta m_N=2$~GeV and $\delta m_+ =480$~MeV.
The sensitivity improves substantially when considering the full spectrum, compared to the line-only case, for masses below $\sim$2 TeV, where the continuum contribution dominates over the line at the end of the spectrum. Sensitivity to this continuum is only possible due to the improved CTA effective area down to a few tens of GeV. The improvement is up to a factor 4 for the smallest mass considered in this work. 
For splitting 1, the Higgsino is within the reach of CTA for the thermal mass of 1~TeV and the first two Sommerfeld-induced resonances.  
In the case of splitting 2, the thermal mass is within reach thanks to the continuum contribution. The first resonance can be probed; the second one only barely.
An accurate calculation of the endpoint contribution, for example using the methods described in~\cite{Beneke:2018ssm,Beneke:2019vhz}, could additionally improve the sensitivity.
The 5$\sigma$ sensitivity is plotted in the upper panels of Fig.~\ref{fig:Higgsinocores}.

We note that despite of the importance of the continuum contribution to CTA's sensitivity, this scenario is not excluded by existing lower energy constraints from {\it Fermi}.
{\it Fermi} observations of Milky Way satellite galaxies constrain the cross-section for annihilation of 1~TeV DM to a $W^+W^-$ final state to be smaller than $\sim 2.7 \times 10^{-25}~{\rm cm^3/s}$.
This is more than an order of magnitude larger than the predicted $W^+W^-$ cross section of the thermal Higgsino, which is $\sim 8.7~(7.7) \times 10^{-27}$ for mass splitting 1 (2) considered above.
In fact, even for the thermal Wino the {\it Fermi} constraint is still a factor of $\sim 2$ from the theoretical prediction, a scenario CTA can probe by several orders of magnitude.

The impact of a cored DM density in the inner Galactic halo is computed for Higgsino DM. Fig.~\ref{fig:Higgsinocores} shows the  expected 95\% C.L. upper limits on $\langle\sigma v\rangle_{\rm line}$ for the Einasto profile (red solid line) and cored profiles with core radii from 300~pc up to 5 kpc.
The expected limits are computed for the two choices of splittings. The degradation of the sensitivity with the increasing size of the DM core follows the same behavior as for the Wino case, since the same DM spatial distribution is assumed for the Wino and Higgsino cases.

As for the Wino, we can also interpret our fiducial limits in the context of a limit on the fraction of DM made up of Higgsinos.
At the thermal 1~TeV value, that fraction is 52\% and 58\% for splitting 1 and 2, respectively.
If we reduced the mass to 0.6~TeV, then a thermally produced Higgsino would only amount to $\sim$36\% of DM~\cite{Cirelli:2007xd}.
Reinterpreting our limits at this mass as a constraint on the DM fraction, we have 36\% and 38\% for splitting 1 and 2, so that the scenario would only marginally be probed.

\subsection{Impact of the Astrophysical Backgrounds}
\begin{figure*}[!htbp] 
\begin{center}
\includegraphics[width=0.45\textwidth]{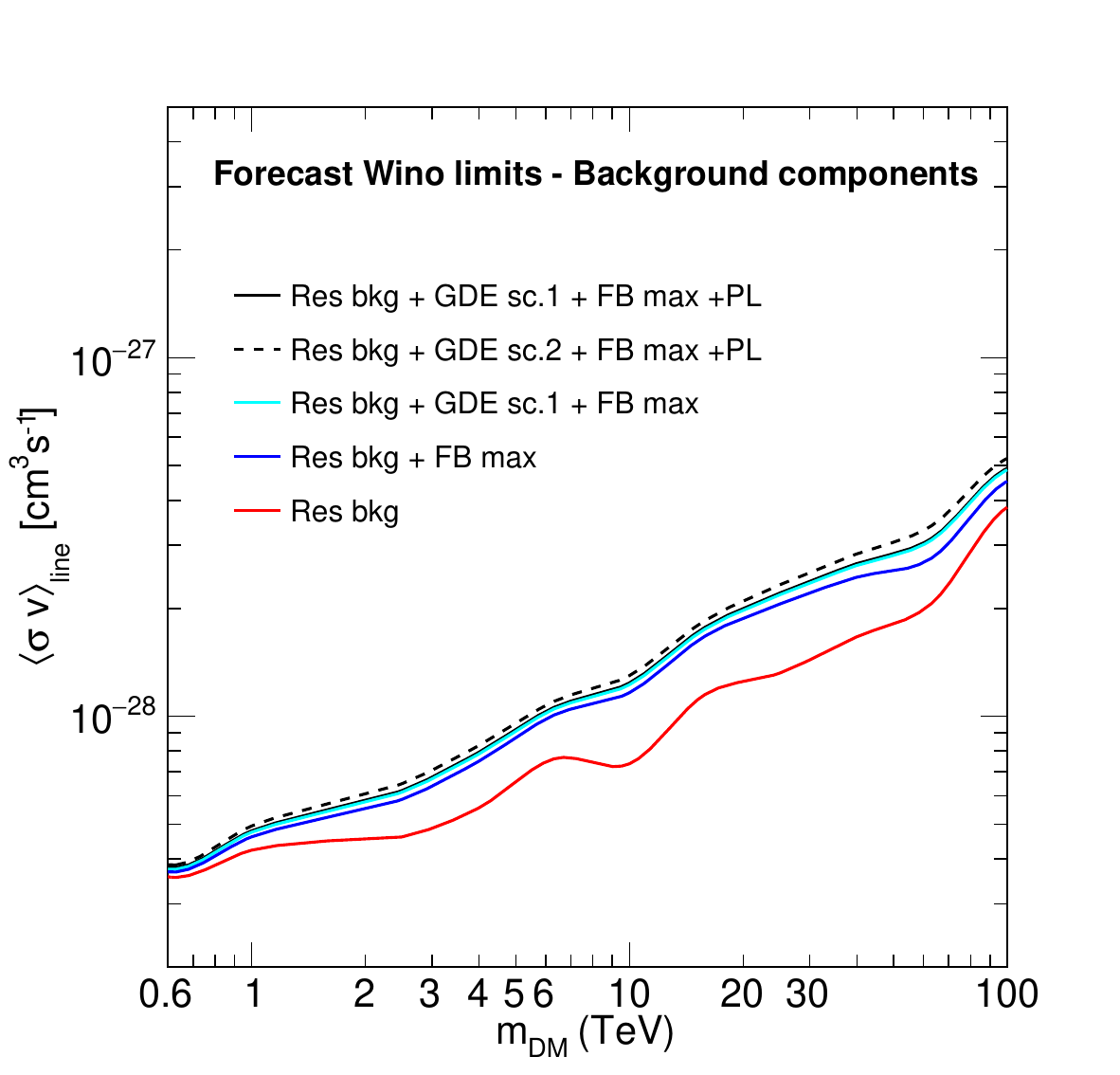}
\includegraphics[width=0.45\textwidth]{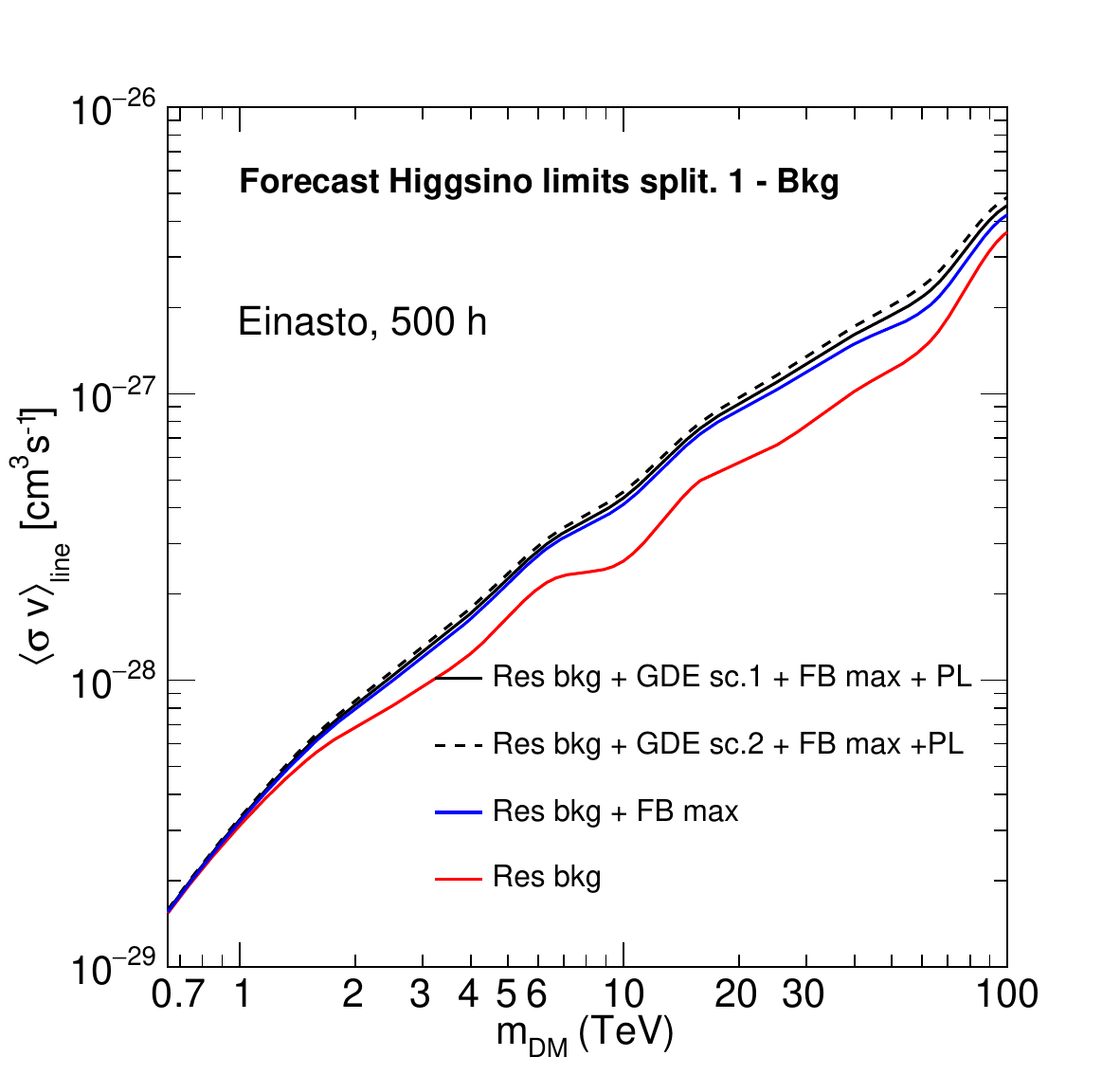}
\caption{Impact on the 95\% C.L. expected mean upper limits, for the Wino and Higgsino annihilation cross sections, of the astrophysical sources of gamma-ray emission that contribute to the overall background. The limits are shown for residual background only (red solid line) and with additional Fermi Bubbles (FB) emission (blue solid line), GDE gamma rays (cyan solid line) and VHE {\it Fermi}-LAT point-like (PL) sources (black solid line). The most impactful background component is the Fermi Bubbles, assuming the ``FB max" spectrum. The point-like (PL) sources have a negligible effect. The impact on the constraints for  the  Wino DM  is given in the left panel, and for Higgsino DM in the right panel assuming splittings $\delta m_N=200$~keV and $\delta m_+=350$~MeV.}
\label{fig:WinoBkgSignalLimits}
\end{center}
\end{figure*}

Fig.~\ref{fig:WinoBkgSignalLimits} shows the impact of each background component on the CTA sensitivity including the full (line + endpoint + continuum) spectrum, for both Wino and Higgsino DM. 
We first show the mean expected 95\% C.L. upper limits assuming a background model that includes all the components previously described, {\it i.e} the residual and all the standard astrophysical components.  Then, we exclude the point-like high-energy {\it Fermi}-LAT sources, the GDE, and eventually the Fermi Bubbles, keeping only the residual background in the last step. This procedure quantifies the impact of each background component on the CTA sensitivity to DM signals.

The limit shown as a black line is computed assuming all the background components given in Eq.~(\ref{eq:bkg}). Masking  
the {\it Fermi}-LAT sources in the overall ROI has a negligible impact on the sensitivity (cyan curve) since the overall solid angle of the masks is negligible with respect to the signal region. 
The blue curve shows the impact of the GDE scenario 1 on the CTA sensitivity. The GDE has an impact up to between 5\% and 20\% for scenario 1, and up to 25\% for scenario 2, for which more emission is expected in the TeV energy range.   

The strongest impact amongst the standard astrophysical backgrounds is due to the Fermi Bubbles assuming the ``FB max'' model. For the Wino, compared to the residual-background-only limits, the ``FB max" model emission has an impact of up to 30\% on the constraints, while for the ``FB min'' model the impact is negligible. The effect of including all standard astrophysical backgrounds, with respect to the residual-background-only case, is up to about 50\% and is largest for masses around ten TeV. 

The impact of the astrophysical background components on the sensitivity to Higgsino DM is shown in the right panel of Fig.~\ref{fig:WinoBkgSignalLimits}, with similar results. A similar degradation is expected for the Wino and Higgsino cases when adding the GDE and Fermi Bubble emissions to the residual background. The observed impact of the astrophysical background on the sensitivity to DM is the same for the Wino and Higgsino candidates. 
 
\begin{figure}[!htbp] 
\begin{center}
\includegraphics[width=0.45\textwidth]{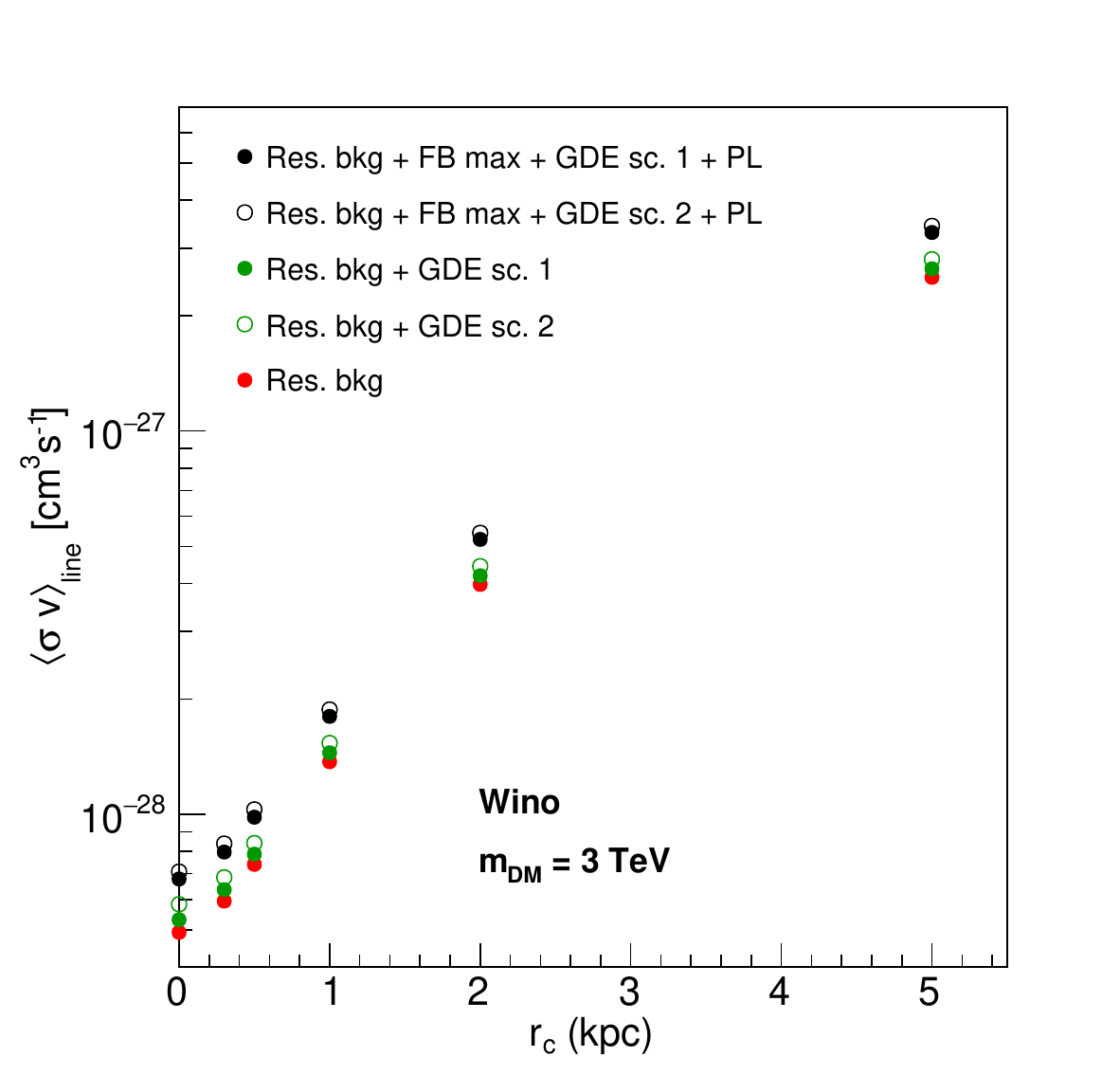}
\caption{95\% C.L. expected mean upper limits on the Wino annihilation line cross section as a function of the DM core radius for a 3 TeV Wino. The limits are shown for the case of residual background only (red dots), for the case of additional astrophysical background from the GDE emission (green dots) in scenarios 1 (green filled dots) and 2 (green empty dots), and for all backgrounds  (black dots).}
\label{fig:GDEmodelling}
\end{center}
\end{figure}

An additional study was performed in order to evaluate the interplay between the spatial morphology of the background components and the size of the DM core. 
Changes in the DM core size are expected to modify which spatial regions have the greatest sensitivity to the DM signal. Consequently, the degree to which the limits vary with core size could change depending on the spatial morphology of the background. Equivalently, the degree to which the astrophysical backgrounds weaken the limits may depend on the assumed core size. For example, background features occurring a few degrees from the GC might have a negligible effect for peaked density profiles with small cores, but a larger impact for few-kpc cores. 
Fig.~\ref{fig:GDEmodelling} shows the CTA expected mean 95\% C.L. upper limit on $\langle\sigma v\rangle_{\rm line}$ for Wino DM, at the thermal mass, as a function of the DM core radius $r_{\rm c}$. The limits are obtained under the assumption of residual background only, residual background and GDE, and all astrophysical backgrounds.
In case of the GDE, the two above-mentioned scenarios are considered.
Note we are considering the case of residual background and GDE, in order to avoid being dominated by the optimistic extrapolation of the Fermi Bubbles in the TeV energy range. We observe no obvious interplay between the core size and the effects of including astrophysical backgrounds; the degradation of the limits in the presence of astrophysical backgrounds is similar for all core sizes tested.)

The sensitivities computed in this work can be significantly degraded when systematic uncertainties are considered in the analysis. DM searches in the complex environment of the GC region will need to contend with experimental systematic uncertainties arising, for instance, from instrumental and observational conditions. Systematic uncertainties will likely dominate the statistical uncertainties, given the large amount of data expected in the GC region. For estimates of the impact of the systematic uncertainties on the sensitivity, see, for instance, Refs.~\cite{Silverwood:2014yza,Lefranc:2015pza,Moulin:2019oyc}.

\section{Conclusions}
\label{sec:conclusion} 
We have computed the CTA sensitivity in the framework of specific heavy DM candidates in the mass range between 600~GeV and 100~TeV, assuming an Einasto DM density profile peaked at the GC and using the most-up-to-date EFT computation of the annihilation spectrum of the Wino and a Sommerfeld-enhanced tree-level computation of the Higgsino spectrum. In this case, CTA has the sensitivity to access to the Wino parameter region in the few tens of TeV mass range, much beyond the thermal mass region,
extending further the already strong constraints by H.E.S.S., and to probe the thermal mass and cross section for the Higgsino. Accordingly, CTA will open unique discovery space for these DM scenarios.

For the Higgsino, we find that the continuum contribution can dominate the line contribution for determining the forecast limits at lower masses (below 2-3 TeV); for the Wino, the inclusion of the endpoint spectrum significantly improves the constraints relative to the case with only the gamma-ray line (as was shown to be the case for H.E.S.S in Ref.~\cite{Rinchiuso:2018ajn}), but adding the continuum contribution does not modify the constraints significantly (see Fig.~\ref{fig:endpointcontribution}).
Given the impressive reach of CTA, the need to calculate this contribution in the case of the Higgsino is clearly emphasised.

As expected, the choice of the DM profile plays an important role in the estimation of the sensitivity in the GC. While for Wino DM, a broad range of masses are within reach even for DM profiles with kpc-size cores, this is not the case for the Higgsino outside the resonances in the predicted annihilation cross section. 

Studying the impact of several standard astrophysical backgrounds on the sensitivity of CTA to DM underlines the importance of performing dedicated studies to both spatially and spectrally model this emission; we estimate that these backgrounds can cause the constraints on the DM annihilation cross section to deteriorate by up to 50\%. In addition, this study showed that the effect of including spatially inhomogeneous astrophysical backgrounds appears to be largely independent of the assumed core scale in the DM density profile.

CTA will be a unique probe for heavy DM candidates in the TeV mass range, improving significantly over the present limits set by the current imaging atmospheric Cherenkov telescopes, although its full impact will depend on the DM candidate and the distribution of the DM around the GC.


\begin{acknowledgments}
This research has made use of the CTA instrument response functions provided by the CTA Consortium and Observatory (version prod3b-v1), see \href{http://www.cta-observatory.org/science/cta-performance/}{cta-observatory.org/science/cta-performance/} for more details.
We are grateful to Matt Baumgart and Varun Vaidya for useful conversations, particularly related to their earlier Higgsino calculation, and we further thank Matt for providing us with the Sommerfeld enhancement for the higher masses considered in this work.
NLR is supported by the Miller Institute for Basic Research in Science at the University of California, Berkeley. OM acknowledges support by World Premier International Research Center Initiative (WPI Initiative), MEXT, Japan and by JSPS KAKENHI Grant Numbers JP17H04836, JP18H04340 and JP18H04578. TRS is partially supported by a John N. Bahcall Fellowship, and by the Office of High Energy Physics of the U.S. Department of Energy under Grant No. DE-SC00012567 and DE-SC0013999. This work made use of resources provided by the National Energy Research Scientific Computing Center, a U.S. Department of Energy Office of Science User Facility supported by Contract No. DE-AC02-05CH11231.

\end{acknowledgments}

\bibliography{bibl}

\end{document}